\documentclass[twocolumn,english,superscriptaddress,prl,nofootinbib]{revtex4-1}
\usepackage{lmodern}
\usepackage{lmodern}
\usepackage[T1]{fontenc}
\usepackage[latin9]{inputenc}
\usepackage{color}
\usepackage{babel}
\usepackage{amsmath}
\usepackage{amssymb}
\usepackage{graphicx}
\usepackage[unicode=true,pdfusetitle,
bookmarks=true,bookmarksnumbered=false,bookmarksopen=false,
breaklinks=false,pdfborder={0 0 1},backref=false,colorlinks=true]
{hyperref}
\hypersetup{
	pdfborderstyle=,urlcolor=orange,linkcolor=blue,citecolor=red}

\makeatletter

\usepackage{color}
\usepackage{babel}
\usepackage{units}

\usepackage{tikz}
\usepackage{color}
\usepackage{dsfont}

\usepackage[cal=boondoxo]{mathalfa}
\usepackage{grffile}

\catcode`,\active

\catcode`\,12

\newsavebox{\@brx}
\newcommand{\llangle}[1][]{\savebox{\@brx}{\(\m@th{#1\langle}\)}%
	\mathopen{\copy\@brx\kern-0.5\wd\@brx\usebox{\@brx}}}
\newcommand{\rrangle}[1][]{\savebox{\@brx}{\(\m@th{#1\rangle}\)}%
	\mathclose{\copy\@brx\kern-0.5\wd\@brx\usebox{\@brx}}}

\makeatletter
\def\maketitle{
	\@author@finish
	\title@column\titleblock@produce
	\suppressfloats[t]}
\makeatother

\allowdisplaybreaks

\makeatother

\begin{document}
	\title[]{Unveiling the structure of wide flat minima in neural networks}
	
	
	\author{Carlo Baldassi}
	\affiliation{Artificial Intelligence Lab, Bocconi University, 20136 Milano, Italy}
	
	\author{Clarissa Lauditi}
	\affiliation{Department of Applied Science and Technology, Politecnico di Torino, 10129 Torino, Italy}
	
	\author{Enrico M. Malatesta}
	\affiliation{Artificial Intelligence Lab, Bocconi University, 20136 Milano, Italy}
	
	\author{Gabriele Perugini}
	\affiliation{Artificial Intelligence Lab, Bocconi University, 20136 Milano, Italy}
	
	\author{Riccardo Zecchina}
	\affiliation{Artificial Intelligence Lab, Bocconi University, 20136 Milano, Italy}
	\begin{abstract}
		The success of deep learning has revealed the application potential of neural networks across the sciences and opened up fundamental theoretical problems. 
		In particular, the fact that learning algorithms based on simple variants of gradient methods are able to find near-optimal minima of highly nonconvex loss functions is an unexpected feature of neural networks. 
		Moreover, such algorithms are able to fit the data 
		even in the presence of noise, and yet they have excellent predictive capabilities. Several empirical results have shown a reproducible correlation between the so-called flatness of the minima achieved by the algorithms and the generalization performance. At the same time, statistical physics results have shown that in nonconvex networks a multitude of narrow minima may coexist with a much smaller number of wide flat minima, which generalize well. Here we show that wide flat minima arise as complex extensive structures, from the coalescence of minima around "high-margin" (i.e., locally robust) configurations. Despite being exponentially rare compared to zero-margin ones, high-margin minima tend to concentrate in particular regions. These minima are in turn surrounded by other solutions of smaller and smaller margin, leading to dense regions of solutions over long distances. Our analysis also provides an alternative analytical method for estimating when flat minima appear and when algorithms begin to find solutions, as the number of model parameters varies.
	\end{abstract}
	\maketitle
	Machine learning has undergone a tremendous acceleration thanks to
	the performance of so-called deep networks~\cite{lecun2015deep}. Very complex architectures
	are able to achieve unexpected performance in very different domains,
	from language processing~\cite{NLP_review1} to protein structure prediction~\cite{senior2020improved,jumper2020high}, just to
	name a few recent impressive results. A key aspect that different
	neural network models have in common is the non-convex nature of the
	learning problem. The learning process must be able to converge in
	a very high-dimensional space and in the presence of a huge number
	of local minima of the loss function which measures the error rate
	on the data set. Surprisingly, this goal can be achieved by algorithms
	designed for convex problems with just few adjustments, such as choosing
	highly parameterized architectures, using dynamic regularization techniques,
	and choosing appropriate loss functions~\cite{MehtaForPhysicists}. In practice, neural networks
	with hundreds of millions of variables can be successfully optimized
	by algorithms based on the gradient descent method~\cite{bottou2010large}.
	
	The study of the geometric structure of the minima of the loss function
	is essential for understanding the dynamic phenomena of learning and
	explaining generalization capabilities. Several empirical results
	have shown a reproducible correlation between the so-called flatness
	of the minima achieved by algorithms and generalization performance~\cite{keskar2016large,jiang2019fantastic,dziugaite2}.
	In a sense that needs to be made rigorous, the loss functions of neural
	networks seem to be characterized by the existence of large flat minima
	that are both accessible and well generalizable~\cite{Draxler,LiVisualizing2018,Huang_Visualizations}. Moreover, similar
	minima are found in the case of randomized labels~\cite{ZhangUnderstandingDL} and different data
	sets, suggesting that they are a robust property of the networks.
	
	This scenario is upheld by some recent studies based on statistical
	physics methods~\cite{baldassi2015subdominant,baldassi2020shaping,relu_locent,BeckerZhangLee2020,ZouHuang2021}, which show that in tractable models of non-convex neural
	networks a multitude of minima with poor generalization capabilities
	coexists with a smaller number of wide flat minima, a.k.a. high local
	entropy minima, that generalize close to optimality~\cite{baldassi2015subdominant}. 
	These studies rely on large-deviation methods that give access to the typical
	number of minima surrounded by a very large number of other minima
	at a fixed distance. The analytical results are corroborated by numerical
	studies that confirm the accessibility of wide flat minima by simple
	algorithms that do not try to sample from the dominating set of minima~\cite{unreasoanable}.
	
	\begin{figure}
		\begin{centering}
			\includegraphics[width=0.99\columnwidth]{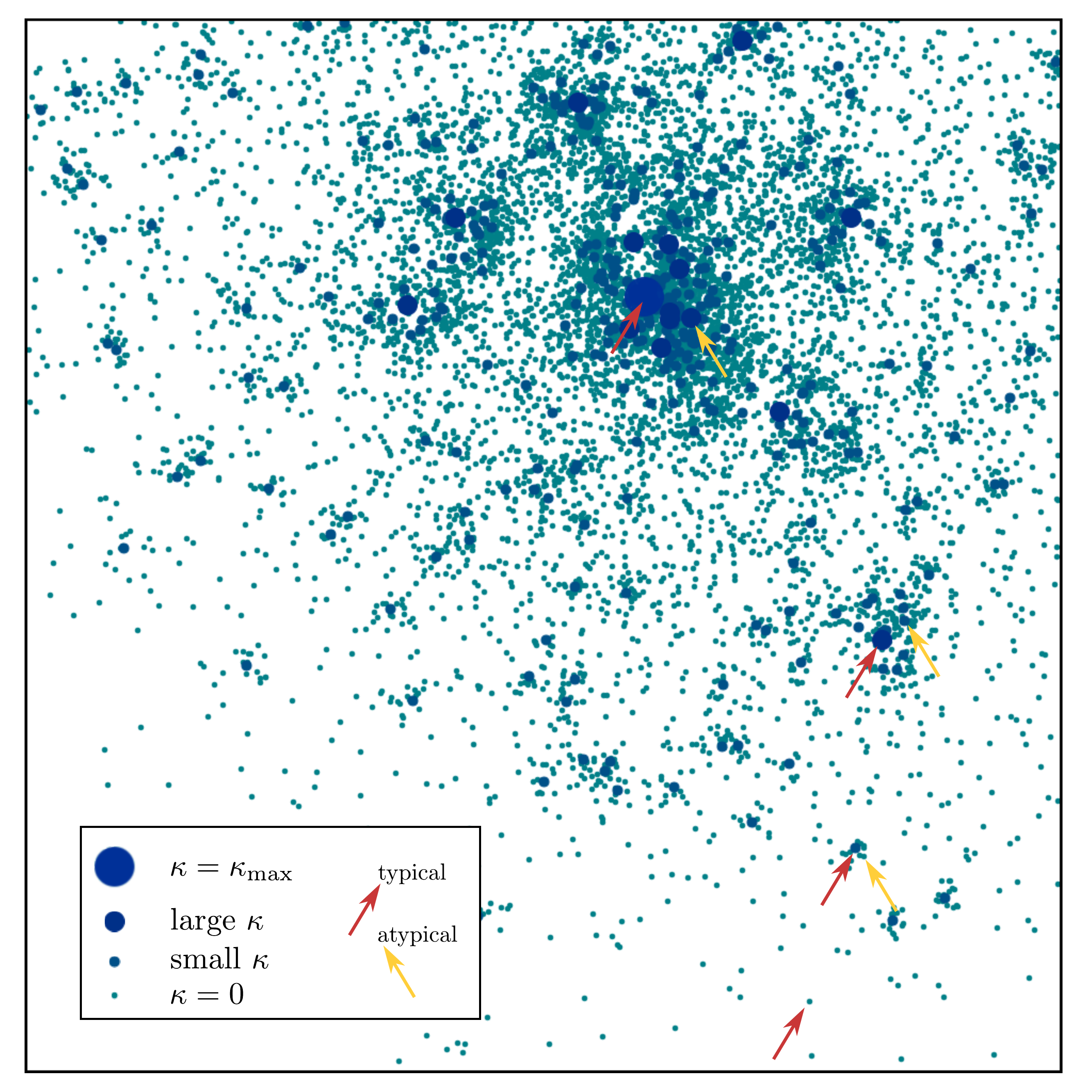}
		\end{centering}
		\caption{
			The picture represents a portion of the space of network configurations. Different dots represent solutions (zero-error configurations); solutions with larger $\kappa$ margin are represented with larger, darker dots (see legend). Red arrows from left to right indicate four examples of typical solutions with a given $\kappa$ (in descending order from top to bottom). The yellow arrows from right to left indicate three examples of the type of atypical solutions found around the typical ones with a larger margin (also in descending order from top to bottom). Low-margin solutions are more numerous than high-margin solutions. Typical low-margin solutions are isolated and distant from each other. Typical high-margin solutions are also distant from each other, but less so, and tend to be surrounded by (atypical) low-margin solutions. Thus, the higher-margin solutions are rare, but they lie in the middle of a dense, extended region that results from the coalescence of the low-margin solutions.
		}
		\label{Fig:sketch} 
	\end{figure}
	
	Here we provide analytical results on the geometric structure of these
	wide flat minima. We take as analytically tractable non-convex model
	a prototypical neural network with $N$ binary weights trained on $P=\alpha N$ random
	patterns, investigated in the thermodynamic limit of large $N$ and large $P$, with
	$\alpha=P/N=O(1)$ 
	. The network performs a binary classification task,
	and its prediction is given by the sign of the output unit. This model has been extensively studied with mean field statistical physics methods~\cite{engel-vandenbroek}, based on the self-averaging property that in the thermodynamic limit the macroscopic behavior of any sample is fully described by the sample average; many of the results were later corroborated by rigorous techniques~\cite{ding2019capacity}. The solutions of the learning task (zero-error configurations) can be characterized by
	their margin, denoted by $\kappa$. The margin
	of a solution is a hard measure of robustness to local perturbations of the
	weights: it is the minimum difference,
	across all the training patterns, between the output pre-activation and
	the threshold. A $\kappa$-margin solution is guaranteed to be surrounded in configuration space by other solutions within a radius proportional to $\kappa \sqrt{N}$. 
	In the model under study, the number of
	solutions at a given margin $\kappa$, when they exist, is typically exponential
	in $N$, i.e. $\exp{\left(N \phi\left(\alpha,\kappa\right)\right)}$. Since $\phi\left(\alpha,\kappa\right)$ is monotonically
	decreasing with $\kappa$, high-margin solutions are exponentially rare
	compared to zero-margin solutions. However, they tend to 
	concentrate in particular regions, and are in turn surrounded by
	other solutions of smaller and smaller margin. This coalescence of minima results in dense regions of solutions over long distances, of size $O\left(N\right)$. 
	This is illustrated in Fig.~\ref{Fig:sketch}, where we show a two-dimensional qualitative sketch of the picture that emerges from our analysis of the geometric distribution of minima for a not too large value of $\alpha$. As as the number of patterns increases (i.e. $\alpha$), the solutions thin out, their margin gets smaller, and above some threshold in $\alpha$ the large connected structures break up and eventually disappear.

	Our results provide a clearer picture regarding the internal structure of the flat minima and allow us to define an alternative analytical method for estimating the threshold at which
	they appear and where the algorithms begin to find solutions efficiently.
	We show that, for values of the loading parameter $\alpha$
	sufficiently small, the zero-error solutions have the following properties:
	
	1) the Hamming distance between typical solutions in the space of network configurations
	is a rapidly decreasing
	function of their margin $\kappa$. Despite being exponentially less numerous (in
	$N$) compared to the $\kappa=0$ solutions, the $\kappa>0$
	solutions tend to have small mutual distance. They are sparser
	and yet much closer.
	2) typical solutions with a prescribed margin $\tilde{\kappa}>0$ are always
	surrounded at $O(N)$ Hamming distance by an exponential number of smaller
	margin solutions. By increasing $\tilde{\kappa}$, we make sure to target
	higher local entropy regions.
	
	While the notion of margin has been developed in the context of shallow
	networks where it can be directly linked to generalization, the notion
	of flatness, or high local entropy, applies also to deep networks
	for which there is no straightforward way to define the margin for the
	hidden layer units. High local entropy minima are stable with respect to
	perturbations of the input and of the internal representations.
	
	\emph{The model.} 
	For simplicity, we discuss here the results of our study by considering  a single-layer~~\cite{Gardner_1989} network with $N$ binary weights $\boldsymbol{w}\in\left\{ -1,1\right\} ^{N}$, which is perhaps the simplest to define non-convex neural network endowed with a non-trivial geometric structure of zero-error solutions. In the SM we detail the analytical results for models with one hidden layer, with binary weights and generic activation functions, which lead to a qualitatively similar geometric scenario. In the SM we also report numerical results for deep networks.
	
	
	Given a (binary) pattern $\boldsymbol{\xi}\in\left\{ -1,1\right\}^{N}$
	as input to the network, the corresponding output is computed as $\sigma_{\text{out}}=\text{sign}\left(\boldsymbol{w}\cdot\boldsymbol{\xi}\right)$.
	We consider a training set composed of $\mu=1,\dots,P=\alpha N$ i.i.d.
	unbiased random binary patterns $\boldsymbol{\xi}^{\mu}=\left\{ -1,1\right\} ^{N}$
	and labels $\sigma^{\mu}=\left\{ -1,1\right\} $~\cite{gardner1988The,gardner1988optimal}.
	The learning problem consists in finding the weights that realize
	all the input-output mappings of the training set. In this paper we
	are interested not only in those configurations that are solutions,
	but also those that have a large confidence level. We quantify
	this by imposing that for every pattern in the training set, the weights
	should have stability $\Delta^{\mu}\equiv\frac{\sigma^{\mu}}{\sqrt{N}}\boldsymbol{w}\cdot\boldsymbol{\xi}^{\mu}$,
	larger than a given margin $\kappa$, which therefore represents the distance from the threshold of the output unit (i.e. the classification boundary) in the direction of the correct label.  The flat measure over these
	configurations is proportional to $\mathbb{X}_{\xi,\sigma}(\boldsymbol{w};\kappa)=\prod_{\mu=1}^{P}\Theta\left(\frac{\sigma^{\mu}}{\sqrt{N}}\sum_{i=1}^{N}w_{i}\xi_{i}^{\mu}-\kappa\right)$
	where $\Theta(\cdot)$ is the Heaviside theta function; this quantity
	is equal to $1$ if the weight $\boldsymbol{w}$ classifies correctly
	all the patterns with a certain margin $\kappa$, and $0$ otherwise.
	The number of solutions with margin $\kappa$ is given by 
	\begin{equation}
	\label{eq::gibbs_measure}
	Z=\sum_{\{w_{i}=\pm1\}}\mathbb{X}_{\xi,\sigma}(\boldsymbol{w};\kappa)
	\end{equation}
	where we have dropped the dependence of $Z$ on $\xi$ and $\sigma$ to lighten the notation. Indeed, $Z$ is the partition function of a flat measure over the $\kappa$-margin solutions, which in turn is the zero-temperature limit of an equilibrium Gibbs measure, with the number of violated patterns as the energy. The corresponding Gibbs entropy of the solutions can thus be obtained as
	\begin{equation}
	\phi(\alpha,\kappa)=\lim\limits _{N\to\infty}\frac{1}{N}\langle\ln Z\rangle_{\xi,\sigma}\label{eq::free_entropy}
	\end{equation}
	where $\langle\dots\rangle_{\xi,\sigma}$ denotes the
	average over random patterns and labels. In the following, we can
	safely impose $\sigma^{\mu}=1$ for every $\mu = 1, \dots, P$ without loss of generality, since
	we can perform the transformation $\xi_{i}^{\mu}\to\sigma^{\mu}\xi_{i}^{\mu}$,
	without affecting the probability measure of the patterns.
	Since we are considering a discrete model, the entropy has a lower bound of $0$.
	In the
	limit of large $N$ the model exhibits a sharp transition
	at the \emph{critical capacity} $\alpha_{c}(\kappa)$, defined as the maximum $\alpha$
	with non-vanishing entropy: $\phi(\alpha_{c}(\kappa),\kappa)=0$.
	For $\alpha<\alpha_{c}(k)$ the probability that an instance of the problem has
	a solution is $1$, but it sharply drops to zero above
	this threshold~\cite{krauth1989storage} (see also~\cite{ding2019capacity}
	for a recent rigorous proof of the value for zero margin $\alpha_{c}(0)\simeq0.833$).
	
	\emph{Distances between typical solutions.} We have computed the entropy
	of solutions, given in equation~\eqref{eq::free_entropy} using the
	replica method. The details of the derivation are given in the Supplemental
	Material (SM). 
	
	\begin{figure}
		\begin{centering}
			\includegraphics[width=0.99\columnwidth]{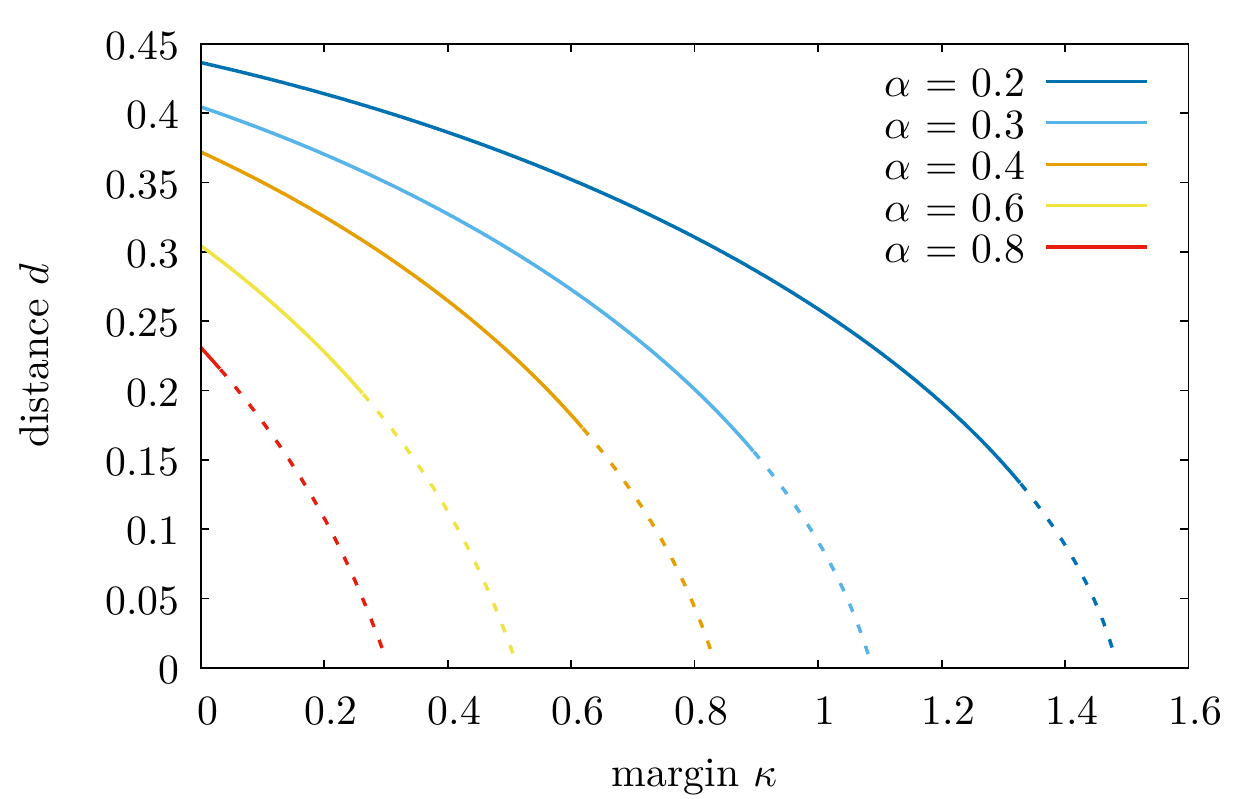}
		\end{centering}
		\caption{Hamming distance between typical solutions as a function
			of the margin imposed, for $\alpha=0.2$, 0.3, 0.4, 0.6 and 0.8 (from
			top to bottom). The lines change from solid to dashed when the entropy
			of solutions becomes negative, i.e. when $\kappa=\kappa_{\text{max}}$
			as defined in the main text.}
		\label{Fig::distance} 
	\end{figure}
	As displayed in Fig.~\ref{Fig::distance}, we find that the Hamming distance
	between solutions is a rapidly decreasing function of the margin.
	As mentioned in the introduction, the entropy is a decreasing function of the
	margin as well (see the SM); this means that even if solutions with larger margin are exponentially
	fewer, they are less dispersed. The closest
	solutions are those with maximum margin $\kappa_{\text{max}}(\alpha)$, defined
	as the largest $\kappa$ with non-vanishing entropy: $\phi(\alpha,\kappa_\mathrm{max}(\alpha))=0$.
	
	\emph{Isolated and and non-isolated solutions.} A key question is how, below the critical capacity, the solutions are
	arranged and how the structure of solution space affects the performance
	of learning algorithms. As discussed by Krauth and Mezard~\cite{krauth1989storage} and Huang and Kabashima~\cite{huang2014origin}
	the structure of typical solutions for $\kappa=0$ consists of clusters
	of vanishing entropy (so called\emph{ frozen}-1RSB scenario). In the
	whole phase below $\alpha_c(\kappa=0)$, zero-margin solutions are isolated, meaning that
	one has to flip an extensive number of weights in order to find the
	closest solution. This scenario was also recently confirmed in simple
	one-hidden layer neural networks with generic activation functions~\cite{relu_locent}
	and also rigorously for the symmetric perceptron~\cite{perkins2021frozen,abbe2021proof}.
	This kind of landscape with point-like solutions suggests that finding
	such solution should be a hard optimization problem; however, this
	is contrary to the numerical evidence given by simple algorithms such
	as the ones based on message passing~\cite{Braunstein2006,baldassi2007efficient}.
	This apparent contradiction was solved in~\cite{baldassi2015subdominant,unreasoanable,baldassi_local_2016}
	where it was shown that there exist rare but dense regions of solutions
	that are accessible by algorithms. Subsequent works suggested that
	simple algorithmic strategies that are commonly used in deep learning
	such as the choice of the loss and the activation function~\cite{baldassi2020shaping,relu_locent}
	or the effect of regularization~\cite{baldassi2020wide} seem to
	help algorithms to access those regions. 
	Finally, a systematic study of the loss landscape of neural networks suggested that as network depth increases the number of minima increase as well, but at the same time they become more clustered and generally are separated by low barriers~\cite{Verpoort2020,BeckerZhangLee2020,ZouHuang2021}. In~\cite{FengTu2021} the authors show that SGD-based algorithms are able to access flat minima because they intrinsically possess an anisotropic noise that is stronger in the directions where the landscape is rough and smaller when it is flat.
	
	\begin{figure}
		\begin{centering}
			\includegraphics[width=0.98\columnwidth]{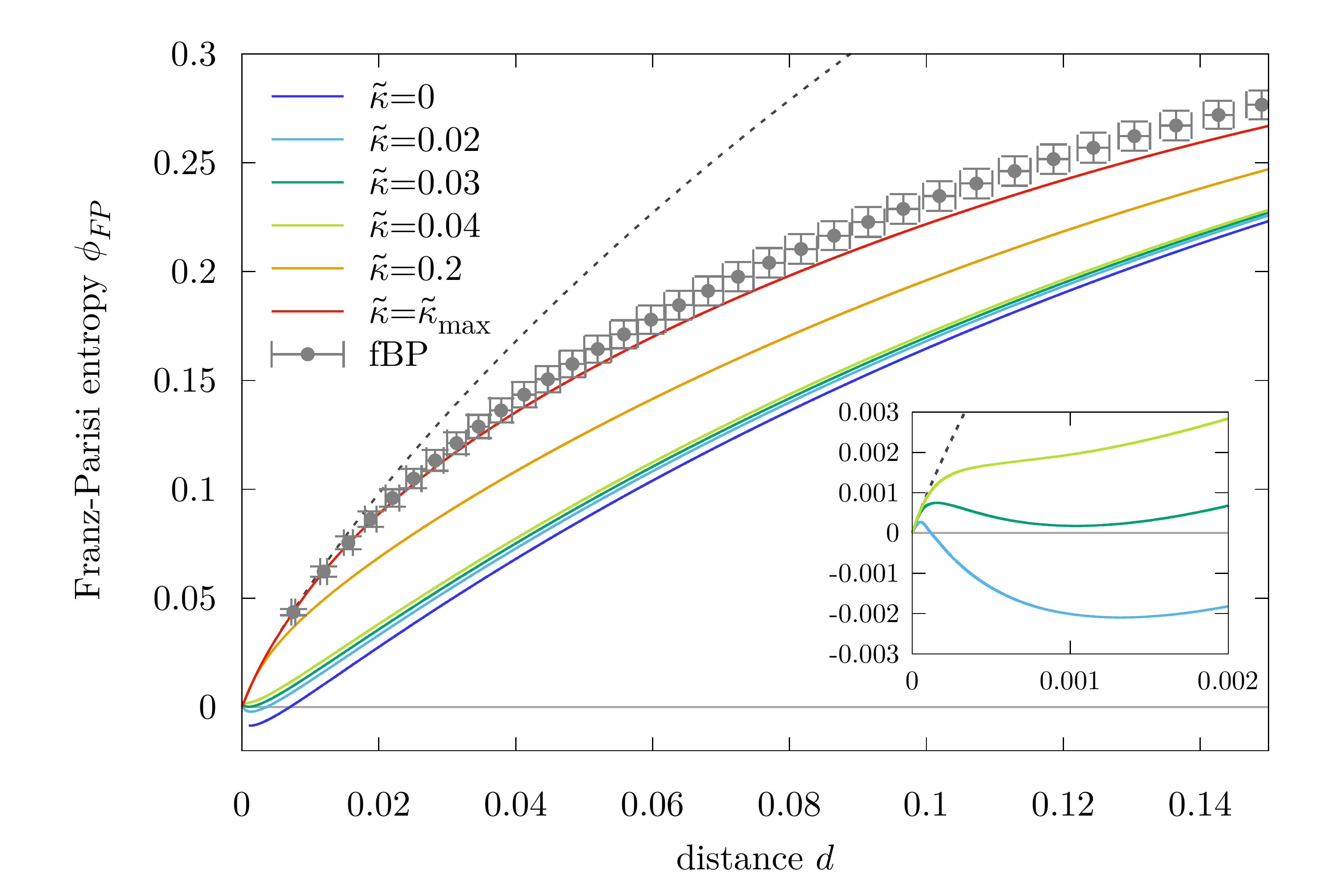}
		\end{centering}
		\caption{Local entropy profiles (with zero margin $\kappa=0$) of typical solutions at $\alpha=0.5$ as a function of the distance, for various values of $\tilde{\kappa}$. The dashed line is the geometrical upper bound obtained by counting all the configurations, corresponding to the unconstrained system with $\alpha=0$. The inset shows a detail of the three curves for $\tilde{\kappa}=0.02$, $0.03$ and $0.04$ in the small-$d$ range. We observe that for $\tilde{\kappa}=0$ the solutions are isolated (the curve is missing for small distances due to numerical issues, but see \cite{huang2014origin}). For $0<\tilde{\kappa}<\tilde{\kappa}_{\min}\simeq0.03$ the entropy has a small positive dense region at small $d$ and there is an interval where it is negative (see the $\tilde{\kappa}=0.02$ curve). For  $\tilde{\kappa}_{\min}\le\tilde{\kappa}<\tilde{\kappa}_u\simeq0.04$ the profiles are all positive, but there are two maxima (see the $\tilde{\kappa}=0.03$ curve). For larger $\tilde{\kappa}$, they grow monotonically up to the global maximum located at a comparatively large distance $d^*(\tilde{\kappa})$ (not visible). The entropy is a monotonic function of $\tilde{\kappa}$ for all distances up to $d^*(\tilde{\kappa}_{\max})\simeq0.285$, and the highest curve is the one for $\tilde{\kappa}_{\max}\simeq0.418$. The points with error bars show the results of numerical experiments (10 samples at $N=2001$ obtained with the focusing-BP algorithm, local entropy estimated by Belief Propagation, see the SM). By design, the fBP algorithm finds solutions in high-local-entropy regions.}
		\label{Fig::FPentropy} 
	\end{figure}
	
	Here we want to better understand the geometry of those rare dense
	regions, in particular how they relate to the $\kappa>0$ solutions, with
	which they share at least the property of being robust with respect to input perturbations
	(see~\cite{relu_locent} for a discussion of the distribution of the stabilities inside a high-local-entropy region). To this end we begin by analyzing in which part of the landscape
	high-margin solutions tend to be concentrated. Given a configuration
	$\tilde{\boldsymbol{w}}$, that we also call the ``reference'',
	we define the \emph{local entropy} of $\tilde{\boldsymbol{w}}$ as
	the logarithm (divided by $N$) of the quantity:
	\begin{equation}
	\mathcal{N}_\xi(\boldsymbol{\tilde{w}},d,\kappa)=\sum_{\boldsymbol{w}}\mathbb{X}_{\xi}(\boldsymbol{w};\kappa)\,\delta\left(N(1-2d)-\sum_{i=1}^{N}\tilde{w}_{i}w_{i}\right)\,.
	\end{equation}
	This expression counts the number of configurations $\boldsymbol{w}$ that are solutions
	with margin $\kappa$ of the classification task, and which lay at a normalized Hamming
	distance $d$ from the reference $\tilde{\boldsymbol{w}}$. Studying the local entropy profile as we vary the distance $d$ thus allows to characterize the density of solutions (with given $\kappa$) in an extensive neighborhood of any given configuration. We are interested in describing the surroundings of typical solutions of given margin $\tilde{\kappa}$, as sampled from the Gibbs measure eq.~\eqref{eq::gibbs_measure}. Thanks to the self-averaging property, for sufficiently large $N$ the local entropy profile is completely captured by the average over the choice of the reference solution and the training set, i.e. by the so-called Franz-Parisi potential~\cite{franz1995recipes,huang2014origin}:
	\begin{equation}
	\phi_{FP}(d;\alpha,\tilde{\kappa},\kappa)=\frac{1}{N}\left\langle \,\frac{1}{Z}\sum_{\tilde{\boldsymbol{w}}}\mathbb{X}_{\xi}(\tilde{\boldsymbol{w}};\tilde{\kappa})\ln\mathcal{N}_\xi(\tilde{\boldsymbol{w}},d,\kappa)\right\rangle _{\xi}\,.
	\end{equation}
	This quantity can be computed with the Laplace method; we report in the Supplemental Material (SM) the technical details of the
	computation, which we performed within the so-called Replica Symmetric
	ansatz for the order parameters. Within this ansatz, finite negative entropies may appear, signalling that the ansatz is incorrect and that the true number of solutions $\mathcal{N}_\xi$ is $0$~\cite{krauth1989storage}.
	We found that,
	for any value of $\alpha$ in the range $0<\alpha<\alpha_{c}(\kappa)$, there
	are several phases depending on the value of $\tilde{\kappa}$ (shown in Fig.~\ref{Fig::FPentropy}):
	\begin{enumerate}
		\item For $\tilde{\kappa}=0$ we recover the results of Huang and Kabashima~\cite{huang2014origin}:
		the solutions are isolated, meaning that the Franz-Parisi entropy is always
		negative in a neighborhood of $d=0$. Here $\phi_{FP}(d)$ has only one maximum with positive entropy at large distances (this
		maximum is present for all the values of $\tilde{\kappa}$ and
		is located at the typical distance $d_{\text{typ}}$ between
		solutions with margin $\tilde{\kappa}$ and $\kappa$).
		\item When $0<\tilde{\kappa}\le\tilde{\kappa}_{\text{max}}(\alpha)$ there always exists a neighborhood of $d=0$ where the average local entropy
		is positive, meaning that typical solutions with non-zero margin are
		always surrounded by an exponential number of solutions having zero margin.
		Furthermore, for small distances the local entropy is nearly indistinguishable
		from the total number of configurations at that distance: almost all configurations 
		around the reference solution are themselves solutions, 
		up to a small, but still $O(N)$, Hamming distance. This means that the cluster is dense.
		\item There exists a $\tilde{\kappa}_{\text{min}}(\alpha)>0$ such that if
		$0<\tilde{\kappa}<\tilde{\kappa}_{\text{min}}(\alpha)$ the local
		entropy is negative in an interval of distances $d\in[d_{1},d_{2}]$
		not containing the origin. This means that no solutions can be found
		in a spherical shell of radius $d\in[d_{1},d_{2}]$.
		\item There exists a $\tilde{\kappa}_{u}(\alpha)>0$ such that if $\tilde{\kappa}_{\text{min}}(\alpha)<\tilde{\kappa}<\tilde{\kappa}_{u}(\alpha)$
		the local entropy is positive, but it is non monotonic. Notice that
		for $0<\tilde{\kappa}<\tilde{\kappa}_{u}$ the Franz-Parisi entropy
		develops a secondary maximum at short distances. This means that typical solutions with such $\tilde{\kappa}$ are immersed within small regions that have a characteristic size - they can be described as isolated (for $\tilde{\kappa} <\tilde{\kappa}_{\min}$) or ``entropically'' isolated (for $\tilde{\kappa}>\tilde{\kappa}_{\min}$) balls.
		\item When $\tilde{\kappa}>\tilde{\kappa}_{u}(\alpha)$ (which can only happen if $\tilde{\kappa}_u(\alpha)<\tilde{\kappa}_{\max}(\alpha)$) the local entropy is monotonic and there is one global maximum, at large
		distances. This suggests that typical solutions with large enough $\tilde{\kappa}$ are immersed in dense regions that do not seem to have a characteristic size and may extend to very large scales: the high-local-entropy regions. We speculate that this property is related to the accessibility of such regions by algorithms.
	\end{enumerate}
	The picture described in the points above stays qualitatively the
	same if we take $\kappa>0$ and $\tilde{\kappa}\ge\kappa$. In particular
	it is interesting to note that typical solutions with a given margin
	$\tilde{\kappa}$ are isolated with respect to solutions with the
	same margin $\kappa=\tilde{\kappa}$. However typical solutions with
	margin $\tilde{\kappa}$ are always surrounded by an exponential number
	of solutions with lower margin $\kappa<\tilde{\kappa}$. 
	
	Therefore we can conclude that even if high-margin solutions are completely
	isolated from each other, they tend to be closer and to concentrate into the rare high local entropy regions of solutions with
	lower (potentially zero) margin. These results help to unravel the
	structure of regions of high local entropy in neural networks: we
	can see them as the union of the typical isolated configurations having
	non-zero margin $\tilde{\kappa}$; those are in turn surrounded by
	solutions with smaller and smaller margin $\kappa<\tilde{\kappa}$.
	We have also checked the validity of the Replica-Symmetric approximation by considering a more general ansatz for the Laplace computation, i.e. the so called one-step Replica Symmetry Breaking scheme (see the SM for additional details).

	\emph{Dense cluster threshold.}
	\begin{figure}
		\begin{centering}
			\includegraphics[width=0.493\columnwidth]{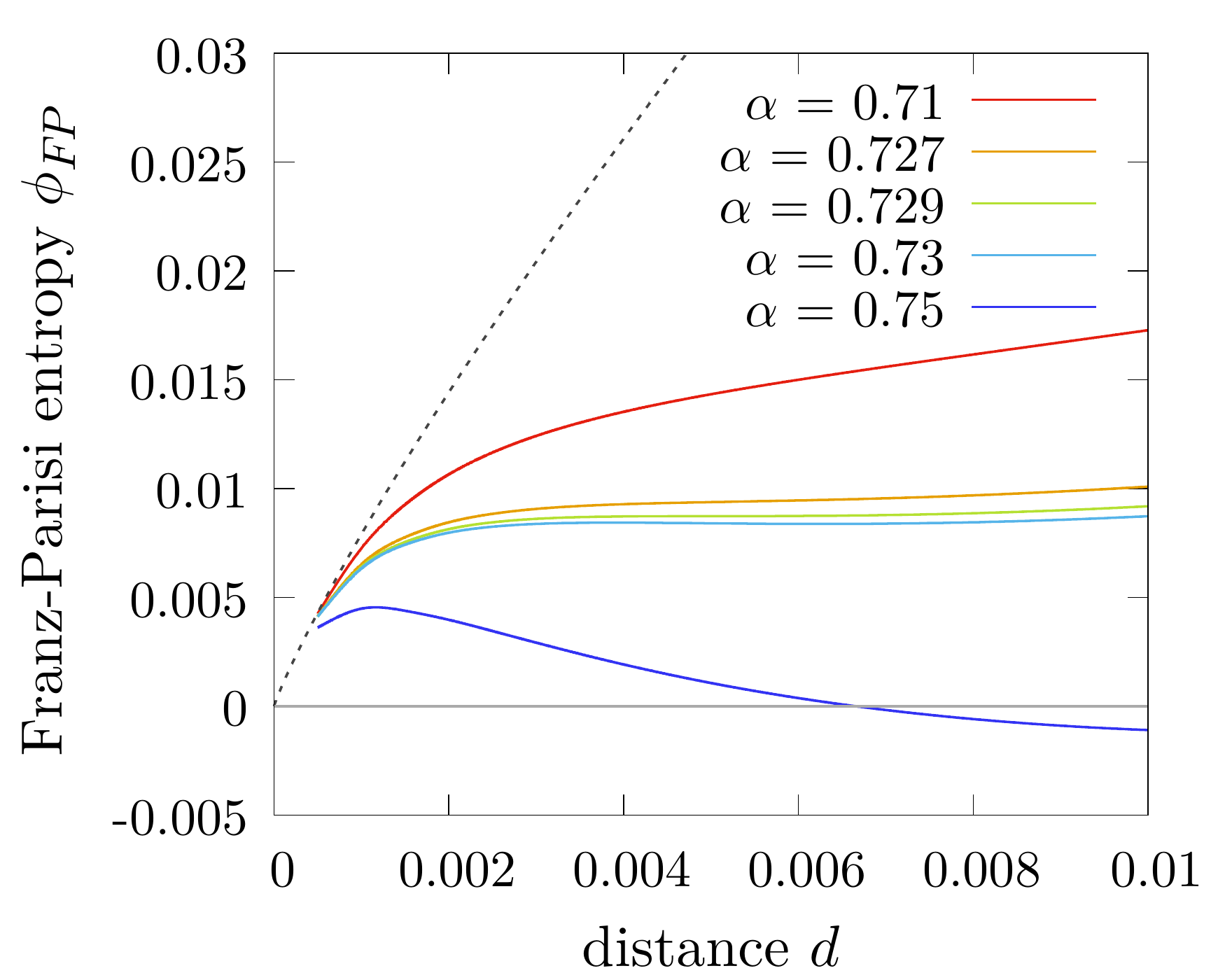}
			\includegraphics[width=0.493\columnwidth]{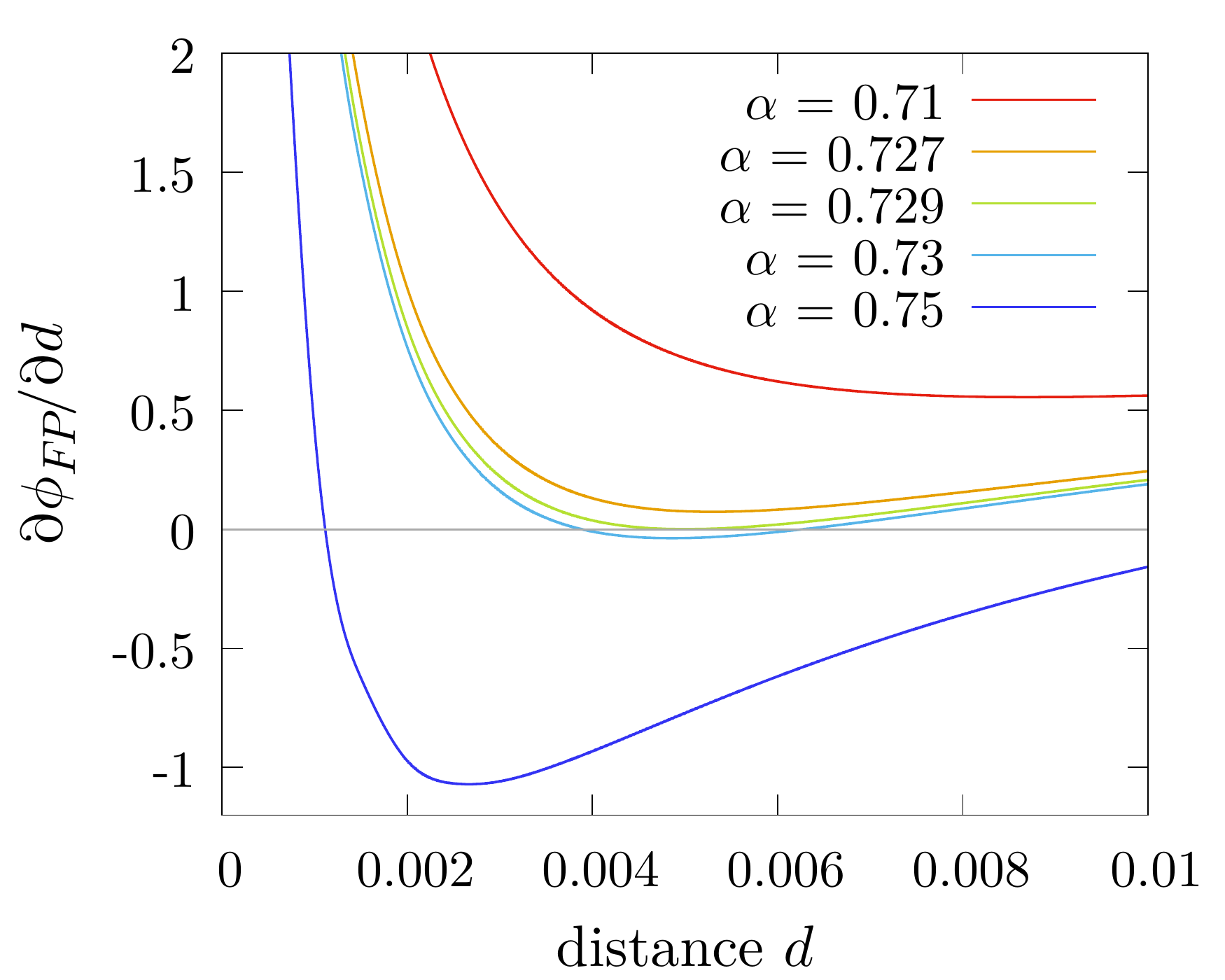}
		\end{centering}
		\caption{Local entropy profiles (with zero margin $\kappa = 0$) of typical maximum margin solutions (left panel) and its derivative (right panel) as a function of the distance. Different values of $\alpha$ are displayed: for $\alpha=0.71$ and $0.727$ the entropy is monotonic, i.e. it has a unique maximum at large distances (not visible). For $\alpha = \alpha_{u}^\prime \simeq 0.729$ the local entropy starts to be non-monotonic (its derivative with respect to the distance develops a new zero). The entropy becomes negative at larger $\alpha$ (i.e. $\tilde{\kappa}_{\max}(\alpha) < \tilde{\kappa}_{\min}(\alpha)$) in a given range of distances.}
		\label{Fig::FP_kmax} 
	\end{figure}
	It has been previously discussed by
	using a large-deviation approach~\cite{baldassi2015subdominant,baldassi_local_2016}
	how the geometrical structure of the high-local-entropy cluster changes with the number
	of patterns $\alpha N$. What was found is that the geometrical structure
	of the cluster remains connected up until a certain value $\alpha_{u}$
	above which the cluster fractures.
	Numerical experiments show also that this geometrical transition strongly
	affects the behavior of algorithms: $\alpha_{u}$ is conjectured to
	be an upper bound for the capacity of efficient learning of algorithms~\cite{unreasoanable}.
	
	As discussed in point 5 of the previous section, a similar situation occurs when considering typical high-margin solutions. Let us define the value $\alpha_{u}^\prime$ as the largest $\alpha$ for which the ``large-scale'' phase exists. It is characterized by the property $\tilde{\kappa}_u(\alpha_{u}^\prime)=\tilde{\kappa}_{\max}(\alpha_{u}^\prime)$. Beyond this value, only the ``isolated balls'' phase (points 3 and 4 in the previous section) remains. Indeed, we found this $\alpha_{u}^\prime$ to be only slightly smaller than the upper bound $\alpha_u$ derived from the large-deviation analysis. Thus, $\alpha_{u}^\prime$ can be used to provide an easier estimate for the algorithmic upper bound.
	
	
	This is illustrated in Fig.~\ref{Fig::FP_kmax}, where we show some plots of $\phi_{FP}(d;\alpha,\tilde{\kappa}_{\text{max}}(\alpha), \kappa)$
	(and its derivative with respect to the distance) for several values of
	$\alpha$. At $\alpha=\alpha_{u}^\prime \simeq 0.73$ the derivative of the local entropy develops a new zero. In this case $\alpha_u \simeq 0.77$.
	
	The discrepancy between the two thresholds can be mainly ascribed to the fact that in the derivation of $\alpha_{u}^\prime$ only typical (albeit high-margin) solutions are considered. (It is also possible that the approximations introduced by the Replica-Symmetric Ansatz play a minor role.) On the other hand, the fact that $\alpha_u \approx \alpha_{u}^\prime$ (which we also checked in an alternative, planted model, the so-called teacher-student scenario) suggests that maximally-dense solutions are not too dissimilar and not too far from maximum-margin solutions. To test this, we performed numerical experiments by sampling solutions found with the focusing-BP algorithm \cite{unreasoanable}, which by design seeks maximally dense solutions, and measured their average local entropy using Belief Propagation (see SM for details). We found that its local entropy profile is only slightly higher than that of the typical $\tilde{\kappa}_{\max}$ solutions, as shown in Fig.~\ref{Fig::FPentropy}.
	This also agrees with previous findings concerning the distribution of stabilities of wide and flat minimizers~\cite{relu_locent} and the impact of certain losses, such as the cross-entropy~\cite{baldassi2020shaping}, which induce a certain degree of robustness during training.

	{\em Discussion and conclusions.} The fracturing transition that sets it  when the curves become non-monotonic is a complex phenomenon. This particular transition was first observed in the aforementioned analysis of large deviations as a transition in $\alpha$. The current scheme also allows us to detect the same transition by observing the space of solutions around typical solutions. In addition, we can also observe a transition in $\tilde k$, where it intersects the value $\tilde k_u(\alpha)$, and a transition in $k$ for fixed $\tilde{\kappa}>\tilde{\kappa}_u\left(\alpha\right)$, (see points 4 and 5 in the previous section). These transitions can be understood as the appearance of a characteristic distance identified by an entropic barrier beyond which the solutions sparsify dramatically. 

	In conclusion we have shown that the dense clusters of solutions which
	are accessed by algorithms in a non-convex model of neural network coincide with
	regions of the weight space where high-margin solutions coalesce.
	While in these regions solutions with the same margin remain mutually isolated,
	they are connected through solutions of smaller margin. These results
	shed light on accessibility and generalization properties, and hopefully
	can help in developing rigorous mathematical results for non-convex neural networks.
	We have verified that similar phenomena take place in one-hidden-layer neural networks with binary weights and generic activation function (we analyzed in particular ReLU and sign activations, see SM Sec.~III) and that numerical results on deeper networks corroborate the scenario (SM Sec.~IV). Also, we refer to the work~\cite{baldassi2021learning} for an analysis on a model with a non-trivial correlated pattern structure, which shows similar qualitative phenomena. 

	\bibliography{references.bib}

	\clearpage

	\title{Unveiling the structure of wide flat minima in neural networks \\ SUPPLEMENTAL MATERIAL}
	\maketitle
	
	\onecolumngrid 

\section{Equilibrium Configurations}

\label{app::Equilibrium} In order to compute the entropy of typical
solutions, we use the replica trick 
\begin{equation}
	\phi=\lim\limits _{N\to\infty}\lim_{n\to0}\frac{\ln\langle Z^{n}\rangle_{\boldsymbol{\xi}}}{nN}
\end{equation}
The replicated partition function can be written as 
\begin{equation}
	\begin{split}Z^{n} & =\int\prod_{a=1}^{N}d\mu(W^{a})\,\prod_{\mu a}\Theta\left(\frac{1}{\sqrt{N}}\sum w_{i}^{a}\xi_{i}^{\mu}-\kappa\right)\\
		& =\int\prod_{a\mu}\frac{dv_{\text{\ensuremath{\mu}}}^{a}d\hat{v}_{\mu}^{a}}{2\pi}\,\prod_{\mu a}\Theta\left(v_{\mu}^{a}-\kappa\right)e^{-iv_{\text{\ensuremath{\mu}}}^{a}\hat{v}_{\mu}^{a}}
		\sum_{\left\{ \boldsymbol{w}^{a}\right\} _{a=1}^{n}}\,e^{i\sum_{\mu,a}\hat{v}_{\mu}^{a}\frac{1}{\sqrt{N}}\sum_{i}w_{i}^{a}\xi_{i}^{\mu}}
	\end{split}
\end{equation}
In the previous equations we have extracted the quantity $v_\mu^a \equiv \frac{1}{\sqrt{N}}\sum w_{i}^{a}\xi_{i}^{\mu}$ using a delta function and its integral representation. 
Now we can perform the disorder average on the patterns in the limit of
large $N$ obtaining 
\begin{equation}
	\begin{split}\prod_{\mu}\left\langle \prod_{i}\,e^{i\frac{\xi_{i}^{\mu}}{\sqrt{N}}\sum_{a}w_{i}^{a}\hat{v}_{\mu}^{a}}\right\rangle _{\boldsymbol{\xi}^{\mu}} & \simeq\prod_{\mu}e^{-\sum_{a<b}\hat{v}_{\mu}^{a}\hat{v}_{\mu}^{b}\left(\frac{1}{N}\sum_{i}w_{i}^{a}w_{i}^{b}\right)-\frac{1}{2}\sum_{a}\left(\hat{v}_{\mu}^{a}\right)^{2}}\,.
	\end{split}
\end{equation}
Next we can introduce the $n \times n$ matrix of order parameters
\begin{equation}
	q_{ab}\equiv\frac{1}{N}\sum_{i=1}^{N}w_{i}^{a}w_{i}^{b}
\end{equation}
which represents the typical overlap between two replicas $a$ and $b$. Correspondingly, we introduce also the matrix of conjugated order parameters
$\hat{q}_{ab}$ (that come from enforcing
the definition of $q_{ab}$ by using delta functions). We finally get
the expression 
\begin{equation}
	\left\langle Z^{n}\right\rangle _{\boldsymbol{\xi}}=\int\prod_{a<b}\frac{dq_{ab}d\hat{q}_{ab}}{(2\pi/N)}\,e^{NS(q_{ab},\,\hat{q}_{ab})}
\end{equation}
where we have defined 
\begin{subequations} 
	\begin{align}
		S(q_{ab},\hat{q}_{ab}) & =-\sum_{a<b}q_{ab}\hat{q}_{ab}+G_{S}(\hat{q}_{ab})+\alpha G_{E}\left(q_{ab}\right)\\
		G_{S} & =\ln\sum_{\left\{ w^{a}\right\} }e^{\sum_{a<b}\hat{q}_{ab}w^{a}w^{b}}\label{Gs}\\
		G_{E} & =\ln\int\prod_{a}\frac{dv^{a}d\hat{v}^{a}}{2\pi}\,\prod_{a}\Theta\left(v^{a}-\kappa\right)e^{i\sum_{a}v^{a}\hat{v}^{a}
			-\frac{1}{2}\sum_{a}\left(\hat{v}^{a}\right)^{2}-\sum_{a<b}q_{ab}\hat{v}^{a}\hat{v}^{b}}%
		\label{Ge}
	\end{align}
\end{subequations}
The entropy is given by 
\begin{equation}
	\phi=\lim\limits _{n\to0}\frac{1}{n}\max\limits _{\left\{ q_{ab},\hat{q}_{ab}\right\} }S(q_{ab},\hat{q}_{ab})\,,
\end{equation}
The values of $q_{ab}$ and $\hat q_{ab}$ are computed by saddle point equations, i.e.
by taking the derivatives of $S$ with respect to $q_{ab}$ and $\hat{q}_{ab}$
and setting them to zero. 

\subsection{Replica-Symmetric Ansatz}

\begin{figure}
	\begin{centering}
		\includegraphics[width=0.493\columnwidth]{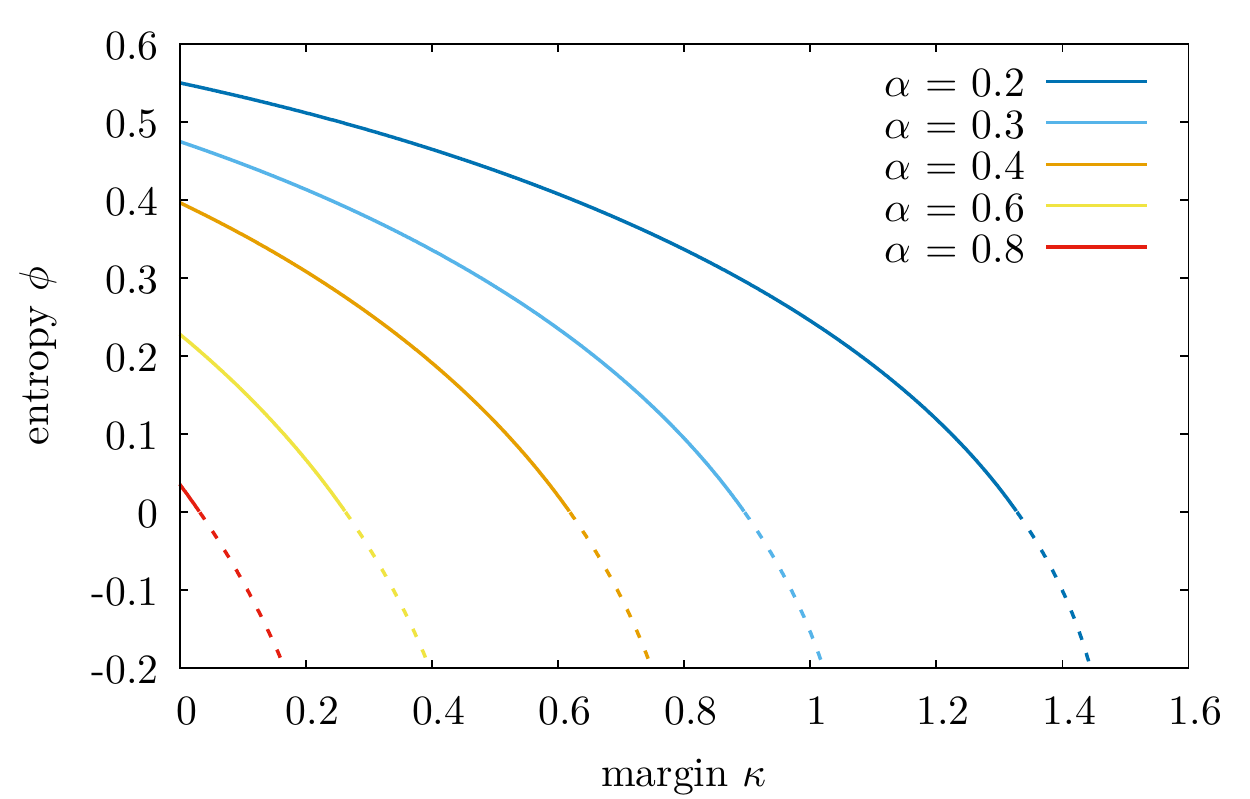}
		\includegraphics[width=0.493\columnwidth]{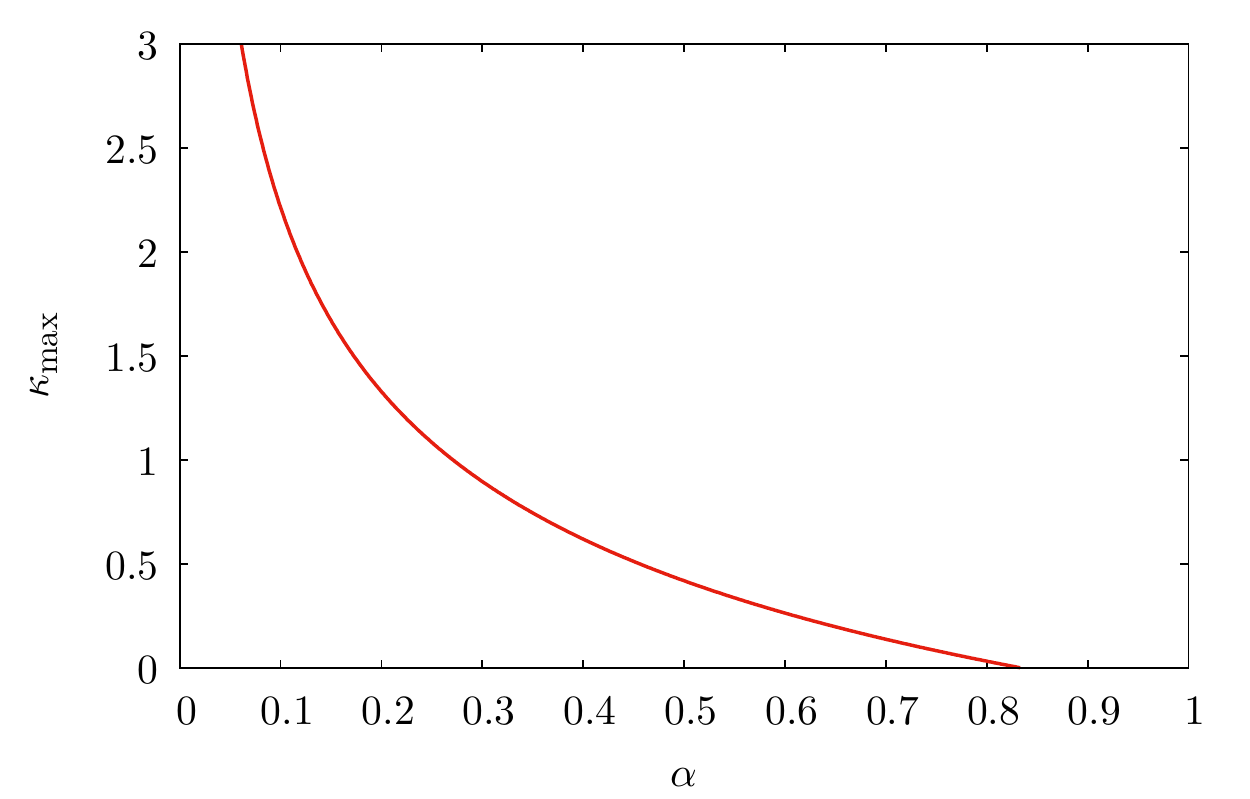}
	\end{centering}
	\caption{Left: RS entropy as a function of the margin $\kappa$ for several values of $\alpha$ and $\kappa_{\text{max}}$ as a function of $\alpha$. Right: the value $\kappa_{\max}$ at which the entropy vanishes, plotted as a function of $\alpha$. This quantity goes to $0$ at $\alpha_c$.}
	\label{Fig::entropy_and_kmax} 
\end{figure}

As investigated by Krauth and M\'ezard~\cite{krauth1989storage},
the most general solution of the saddle point equations (below the
critical capacity of the model) is given by the Replica-Symmetric
(RS) ansatz:
\begin{subequations}
	\begin{align}
		q_{ab} &=\delta_{ab}+q(1-\delta_{ab})\,,\\
		\hat{q}_{ab} &=\hat{q}(1-\delta_{ab})		
	\end{align}
\end{subequations}
The entropy is written as
\begin{equation}
	\phi=-\frac{\hat{q}}{2}(1-q)+\mathcal{G}_{S}+\alpha\mathcal{G}_{E}
\end{equation}
where 
\begin{subequations} 
	\begin{align}
		\mathcal{G}_{S} & \equiv\frac{\hat{q}}{2}+\lim\limits _{n\to0}\frac{G_{S}}{n}=\int Dx\ln2\cosh(\sqrt{\hat{q}}x)\,,\\
		\mathcal{G}_{E} & \equiv\lim\limits _{n\to0}\frac{G_{E}}{n}=\int Dx\ln H\left(\frac{\kappa-\sqrt{q}x}{\sqrt{1-q}}\right)
	\end{align}
\end{subequations}
are called ``entropic'' and ``energetic'' terms, respectively, and $H(x)\equiv\frac{1}{2}\mathrm{erfc}\left(\frac{x}{\sqrt{2}}\right)$. By solving the saddle point equations we can compute the distance between typical solutions by the simple relation $d=\frac{1-q}{2}$, whose plot is reported in Fig.~\ref{Fig::distance} of the main text. We plot in Fig~\ref{Fig::entropy_and_kmax} the entropy as a function of $\kappa$ and for several values of $\alpha$. As discussed in the main text, the maximum margin $\kappa_{\text{max}}$ that we can impose is the one for which the RS entropy vanishes. We plot it in Fig.~\ref{Fig::entropy_and_kmax} as a function of $\alpha$.

\section{The entropy of solutions around reference solutions with different margins.}

\label{app::FP} We briefly review the computation of the Franz-Parisi
entropy~\cite{franz1995recipes} of the perceptron storing random
patterns~\cite{huang2014origin}. Given a configuration $\tilde{\boldsymbol{w}}$
(called ``reference'') we define the local entropy around that configuration,
as the log of $\mathcal{N}_{\xi}(\tilde{\boldsymbol{w}},t_{1})$, i.e. the
number of weights, that are solutions and are with an overlap $t_{1}$
with the reference configuration, i.e. 
\begin{equation}
	\mathcal{N}_{\xi}(\tilde{\boldsymbol{w}},t_{1})=\sum_{\boldsymbol{w}}\mathbb{X}_{\xi}(\boldsymbol{w};\kappa)\,\delta\left(Nt_{1}-\sum_{i=1}^{N}\tilde{w}_{i}w_{i}\right)
\end{equation}
The Franz-Parisi entropy is the average local entropy
around a reference which is a solution extracted from the Gibbs measure:
\begin{equation}
	\phi_{FP}(t_{1})= \frac{1}{N}\left\langle \,\frac{1}{Z}\sum_{\tilde{\boldsymbol{w}}}\mathbb{X}_{\xi}(\tilde{\boldsymbol{w}};\tilde{\kappa})\ln\mathcal{N}_{\xi}(\tilde{\boldsymbol{w}},t_{1})\right\rangle _{\boldsymbol{\xi}}
\end{equation}
To evaluate the $\phi_{FP}(t_{1})$, we use the replica trick twice:
\begin{subequations}
	\begin{align}
		\ln\mathcal{N}_{\xi} & =\lim_{s\to0}\partial_{s}\mathcal{N}_{\xi}^{s}\\
		Z^{-1} & =\lim_{n\to0}Z^{n-1}
	\end{align}
\end{subequations} 
In the following we will use indices $a,b\in[n]$
and $c,d\in[s]$. The Franz-Parisi entropy is:
\begin{equation}
	\begin{split}\phi_{FP}(t_{1}) & =\lim\limits _{\substack{n\to0\\
				s\to0}}\partial_{s}\sum_{\left\{ \tilde{\boldsymbol{w}}^{a}\right\} }\prod_{a}\left\langle \mathbb{X}_{\xi}(\tilde{\boldsymbol{w}}^{a};\tilde{\kappa})\,\mathcal{N}_{\xi}^{s}(\tilde{\boldsymbol{w}}^{a=1},t_{1})\right\rangle _{\boldsymbol{\xi}} =\\
		&= \lim\limits _{\substack{n\to0\\
				s\to0
			}
		}\partial_{s}\int\prod_{\mu a}\frac{dv_{\mu}^{a}d\hat{v}_{\mu}^{a}}{2\pi}\,\Theta\left(v_{\mu}^{a}-\tilde{\kappa}\right)e^{-iv_{\mu}^{a}\hat{v}_{\mu}^{a}}\int\prod_{\mu c}\frac{du_{\mu}^{c}d\hat{u}_{\mu}^{c}}{2\pi}\,\Theta\left(u_{\mu}^{c}-\kappa\right)e^{-iu_{\mu}^{c}\hat{u}_{\mu}^{c}}\\
		& \quad \times\,\sum_{\left\{ \tilde{\boldsymbol{w}}^{a}\right\} }\sum_{\left\{ \boldsymbol{w}^{c}\right\} }\prod_{\mu}\left\langle e^{\frac{i}{\sqrt{N}}\sum_{i}\xi_{i}^{\mu}\left(\sum_{a}\tilde{w}_{i}^{a}\hat{v}_{\mu}^{a}+\sum_{c}w_{i}^{c}\hat{u}_{\mu}^{c}\right)}\right\rangle _{\boldsymbol{\xi}^{\mu}}\prod_{c}\delta\left(Nt_{1}-\sum_{i=1}^{N}\tilde{w}_{i}^{a=1}w_{i}^{c}\right)%
	\end{split}
	\label{eq::FP_replicas}
\end{equation}
where two auxiliary variables have been introduced 
\begin{subequations}
	\label{eq::vu} 
	\begin{align}
		v_{\mu}^{a} & =\frac{1}{\sqrt{N}}\,\sum_{i}\tilde{w}_{i}^{a}\xi_{i}^{\mu}\\
		u_{\mu}^{c} & =\frac{1}{\sqrt{N}}\sum_{i}w_{i}^{c}\xi_{i}^{\mu}
	\end{align}
\end{subequations} 
and we have enforced those definitions by using
delta functions (and their integral representation). Now the average
over patterns can be done in the thermodynamic limit and it reads
\begin{equation}
	\begin{split}\left\langle e^{\frac{i}{\sqrt{N}}\sum_{i}\xi_{i}^{\mu}\left(\sum_{a}\tilde{w}_{i}^{a}\hat{v}_{\mu}^{a}+\sum_{c}w_{i}^{c}\hat{u}_{\mu}^{c}\right)}\right\rangle _{\boldsymbol{\xi}^{\mu}} & \simeq e^{-\frac{1}{2N}\sum_{i}\left(\sum_{a}\tilde{w}_{i}^{a}\hat{v}_{\mu}^{a}+\sum_{c}w_{i}^{c}\hat{u}_{\mu}^{c}\right)^{2}}\\
		& =e^{-\frac{1}{2}\sum_{ab}\left(\frac{1}{N}\sum_{i}\tilde{w}_{i}^{a}\tilde{w}_{i}^{b}\right)\hat{v}_{\mu}^{a}\hat{v}_{\mu}^{b}-\frac{1}{2}\sum_{cd}\left(\frac{1}{N}\sum_{i}w_{i}^{c}w_{i}^{d}\right)\hat{u}_{\mu}^{c}\hat{u}_{\mu}^{d}-\sum_{ac}\left(\frac{1}{N}\sum_{i}\tilde{w}_{i}^{a}w_{i}^{c}\right)\hat{v}_{\mu}^{a}\hat{u}_{\mu}^{c}}
	\end{split}
\end{equation}
The previous expression only depends on first
two moments of the variables~\eqref{eq::vu}. We can therefore define
the order parameters 
\begin{subequations}
	\begin{align}
		q_{ab} &\equiv \left\langle v_{\mu}^{a} v_{\mu}^{b} \right\rangle_{\boldsymbol{\xi}^\mu} 
		= \frac{1}{N} \sum_{i} \tilde{w}^{a}_{i} \tilde{w}^{b}_{i}\,, \\
		p_{cd} &\equiv  \left\langle u^{\mu}_c v^{\mu}_d \right\rangle_{\boldsymbol{\xi}^\mu} = \frac{1}{N} \sum_{i} w^{c}_{i} w^{d}_{i} \,, \\
		t_{ac} &\equiv \left\langle v_{\mu}^{a} u_{\mu}^{c} \right\rangle_{\boldsymbol{\xi}^\mu} 
		= \frac{1}{N} \sum_{i} \tilde{w}^{a}_{i} w^{c}_{i} \,.
	\end{align}
\end{subequations}
Notice that because of the delta function constraining
the reference $\tilde{\boldsymbol{w}}$ and $\boldsymbol{w}$, we
have $t_{1c}=t_{1}$ for every $c$. We can enforce those definitions
again, using delta functions and their integral representations. The
Franz-Parisi entropy can be finally written as 
\begin{equation}
	\begin{aligned}\phi_{FP}(t_{1}) & =\lim\limits _{\substack{n\to0\\
				s\to0
			}
		}\partial_{s}\int\prod_{a<b}\frac{dq_{ab}d\hat{q}_{ab}}{2\pi}\,\prod_{c<d}\frac{dp_{cd}d\hat{p}_{cd}}{2\pi}\,\prod_{c,\,a\ne1}\frac{dt_{ac}d\hat{t}_{ac}}{2\pi}\prod_{c}\frac{d\hat{t}_{1c}}{2\pi}\,e^{NS(q_{ab},\hat{q}_{ab},p_{cd},\hat{p}_{cd},t_{cd},\hat{t}_{cd})}\end{aligned}
	\label{S}
\end{equation}
where we have defined the entropic and energetic terms as: 
\begin{subequations}
	\begin{align}
		S & =-\sum_{a<b}q_{ab}\hat{q}_{ab}-\sum_{c<d}p_{cd}\hat{p}_{cd}-\sum_{ac}t_{ac}\hat{t}_{ac}+G_{S}+\alpha\,G_{E}\\\label{Entropic}
		G_{S} & =\ln\sum_{\left\{ \tilde{w}^{a}\right\} }\sum_{\left\{ w^{c}\right\} }e^{\sum_{a<b}\hat{q}_{ab}\tilde{w}^{a}\tilde{w}^{b}\,+\,\sum_{c<d}\hat{p}_{cd}w^{c}w^{d}\,+\,\sum_{ac}\hat{t}_{ac}\tilde{w}^{a}w^{c}}\\
		\begin{split}G_{E} & =\ln\int\prod_{a}\frac{dv^{a}d\hat{v}^{a}}{2\pi}e^{iv_{a}\hat{v}_{a}}\int\frac{du^{c}d\hat{u}^{c}}{2\pi}e^{iu_{c}\hat{u}_{c}}\prod_{a}\Theta(v_{a}-\tilde{\kappa})\prod_{c}\Theta(u_{c}-\kappa)\,\\ 
			& \quad \times e^{-\frac{1}{2}\sum_{ab}q_{ab}\hat{v}^{a}\hat{v}^{b}-\frac{1}{2}\sum_{cd}p_{cd}\hat{u}^{c}\hat{u}^{d}-\sum_{ac}t_{ac}\hat{v}^{a}\hat{u}^{c}}\label{Energetic}
		\end{split}
	\end{align}
\end{subequations}

\subsection{Replica-Symmetric Solution}

The computation can be carried out in the RS ansatz for the order
parameters: 
\begin{subequations} 
	\begin{align}
		q_{ab} & =\delta_{ab}+q(1-\delta_{ab})\\
		p_{cd} & =\delta_{cd}+p(1-\delta_{cd})\\
		t_{ac} & =t_{1}\,\delta_{a1}\,+\,t_{0}\,(1-\delta_{a1})
	\end{align}
\end{subequations} and similarly for their conjugate quantities
\begin{subequations}
	\begin{align}
		\hat{q}_{ab} & =\hat{q}(1-\delta_{ab})\\
		\hat{p}_{cd} & =\hat{p}(1-\delta_{cd})\\
		\hat{t}_{ac} & =\hat{t}_{1}\,\delta_{a1}\,+\,\hat{t}_{0}\,(1-\delta_{a1})
	\end{align}
\end{subequations} 
The entropy is written as 
\begin{equation}
	\label{eq::FP_entropy}
	\phi_{\text{FP}}=-\frac{\hat{p}}{2}(1-p)+t_{0}\hat{t}_{0}-t_{1}\hat{t}_{1}+\mathcal{G}_{S}+\alpha\mathcal{G}_{E}
\end{equation}
where 
\begin{subequations} 
	\begin{align}
		\mathcal{G}_{S} & \equiv\frac{\hat{p}}{2}+\lim\limits _{\substack{n\to0\\
				s\to0
			}
		}\partial_{s}G_{S}=\int Dx\frac{\sum_{\tilde{w}=\pm1}e^{\sqrt{\hat{q}}\tilde{w}x}\int Dy\ln2\cosh\left(\sqrt{\hat{p}-\frac{\hat{t}_{0}^{2}}{\hat{q}}}y+\frac{\hat{t}_{0}}{\sqrt{\hat{q}}}x+(\hat{t}_{1}-\hat{t}_{0})\tilde{w}\right)}{2\cosh\left(\sqrt{\hat{q}}x\right)}\\
		\mathcal{G}_{E} & \equiv\lim\limits _{\substack{n\to0\\
				s\to0
			}
		}\partial_{s}G_{E}=\int Dx\,\frac{{\displaystyle \int Dy\,H\left(\frac{\sqrt{\gamma}\left(\tilde{\kappa}-\sqrt{q}x\right)-(t_{1}-t_{0})y}{\sqrt{\gamma(1-q)-(t_{1}-t_{0})^{2}}}\right)\ln H\left(\frac{\kappa-\frac{t_{0}}{\sqrt{q}}x-\sqrt{\gamma}y}{\sqrt{1-p}}\right)}}{{\displaystyle H\left(\frac{\tilde{\kappa}-\sqrt{q}x}{\sqrt{1-q}}\right)}}
	\end{align}
\end{subequations} 
where we have defined the quantity $\gamma\equiv p-\frac{t_{0}^{2}}{q}$.

\subsection{One-step Replica Symmetry Breaking Solution}

As reported in the main text we have tested the goodness of the RS approximation by plugging a more general ansatz for the order parameters. Since we have not changed the structure of the order parameters of the reference, we have imposed a one-step replica symmetry breaking ansatz (1RSB) for the order parameters that only involve the constrained configuration (i.e. is $p_{cd}$ and $\hat p_{cd}$), leaving $q_{ab}$, $t_{ac}$ and their conjugated parameters unchanged. The new ansatz for $p_{cd}$ and $\hat p_{cd}$ is
\begin{subequations} 
	\begin{align}
		p_{cd} &= p_0 + (p_1-p_0) I_{cd}^{(n,m)} + (1-p_1) I_{cd}^{(n,1)}\\
		\hat p_{cd} &= \hat p_0 + (\hat p_1 - \hat p_0) I_{cd}^{(n,m)} + (1-\hat p_1) I_{cd}^{(n,1)}
	\end{align}
\end{subequations}
where $I_{cd}^{(n,m)}$ is the $(c,d)$ element of a block matrix of size $n \times n$ whose diagonal blocks have size $m \times m$ and contain all ones and outside of them the matrix is composed of zeros
Therefore $p_1$ represents the overlap between different constrained solutions belonging to the same block while $p_0$ characterizes the overlap between two solutions belonging to different blocks of replicas. 

In this framework, it is convenient to rewrite the expression of $G_S \left( \hat{q}_{ab}, \hat{p}_{cd}, \hat{t}_{ac}\right)$ and $G_{E} \left(q_{ab}, p_{cd}, t_{ac} \right)$ given respectively in equations~\eqref{Entropic} and~\eqref{Energetic} as
\begin{subequations}
	\begin{align}
		G_{S} &=\ln\sum_{\left\{ \tilde{w}^{a}\right\} }\sum_{\left\{ w^{c}\right\} }e^{\frac{\hat{q}-\hat{t}_0}{2}\left( \sum_{a} \tilde{w}^a\right)^2 + \frac{\hat{p}_0 - \hat{t}_0}{2} \left(\sum_{c} w^{c} \right)^2 + \frac{\hat{t}_0}{2}\left( \sum_{a} \tilde{w}^a + \sum_{c} w^{c}\right)^2 + \frac{\hat{p}_1 - \hat{p}_0}{2}\sum_{i=1}^{s/m}\left(\sum_{c_i = 1}^{m} w^{c_i} \right)^2 + (\hat{t}_1 - \hat{t}_0 ) \tilde{w}^1 \sum_{c} w^{c} - \frac{\hat{q}}{2}n - \frac{\hat{p}_1}{2}s}\\
		\begin{split}G_{E} & =\ln\int\prod_{a}\frac{dv^{a}d\hat{v}^{a}}{2\pi}\,e^{i \sum_{a} v_{a}\hat{v}_{a}}\int\prod_{c}\frac{du^{c}d\hat{u}^{c}}{2\pi}\,e^{i\sum_{c}u_{c}\hat{u}_{c}}\prod_{a}\Theta(v^{a}-\tilde{\kappa})\prod_{c}\Theta(u^{c}-\kappa)\,\\ 
			& \times e^{-\frac{q -t_0}{2}\left( \sum_{a} \hat{v}^{a}\right)^2 - \frac{p_0 - t_0}{2}\left( \sum_{c} \hat{u}^c\right)^2 - \frac{t_0}{2}\left( \sum_{a} \hat{v}^a + \sum_{c} \hat{u}^c\right)^2 + (t_0 - t_1) \,\hat{v}^1 \sum_{c} \hat{u}^c - \frac{p_1 - p_0}{2}\sum_{i} \left(\sum_{c_i} \hat{u}^{c_i} \right)^2}\\
			& \times e^{- \frac{1-q}{2}\sum_{a} \left(\hat{v}^a\right)^2 - \frac{1- p_1}{2}\sum_{c} \left( \hat{u}^c\right)^2}
		\end{split}
	\end{align}
\end{subequations}
where the replica index $c = 1, \dots, s$ is rewritten as $c = (i-1) m + c_i$ where $i = 1, \ldots , s/m$ is the block index $c$ belongs to and $c_i = 1, \ldots , m$ specifies the position inside the block. 

Using appropriate Hubbard-Stratonovich transformations and changing variables to perform some integrals explicitly, we get to the final expression for the Franz-Parisi entropy
\begin{equation}
	\phi_{FP}(t_1) = - t_1 \hat{t}_1 + t_0 \hat{t}_0 - \frac{m}{2}\left(p_1 \hat{p}_1 - p_0 \hat{p}_0 \right) - \frac{\hat{p}_1}{2} \left(1- p_1 \right) + \mathcal{G}_{S} + \alpha \mathcal{G}_{E}
\end{equation}
where the new entropic and energetic terms are
\begin{subequations} 
	\begin{align}
		\mathcal{G}_{S} & \equiv\frac{\hat{p}_1}{2}+\lim\limits _{\substack{n\to0\\
				s\to0
			}
		}\partial_{s}G_{S} \\
		&\nonumber = \frac{1}{m}\int Dx  \,\frac{\sum_{\tilde{w}=\pm1}e^{\sqrt{\hat{q}}\tilde{w}x}\int Dy\ln \int Dz \, \left(2\cosh\left(\sqrt{\hat{p}_0-\frac{\hat{t}_{0}^{2}}{\hat{q}}}y+\frac{\hat{t}_{0}}{\sqrt{\hat{q}}}x + \sqrt{\hat{p}_1 - \hat{p}_0} \, z+(\hat{t}_{1}-\hat{t}_{0})\tilde{w}\right)\right)^m}{2\cosh\left(\sqrt{\hat{q}}x\right)}\\
		\mathcal{G}_{E} & \equiv\lim\limits _{\substack{n\to0\\
				s\to0
			}
		}\partial_{s}G_{E} \\
		&\nonumber=\frac{1}{m}\int Dx\,\frac{{\displaystyle \int Dy\,H\left(\frac{\sqrt{\gamma}\left(\tilde{\kappa}-\sqrt{q}x\right)-(t_{1}-t_{0})y}{\sqrt{\gamma(1-q)-(t_{1}-t_{0})^{2}}}\right)\ln \int Dz \,  H^{m}\left(\frac{\kappa - \sqrt{p_1 - p_0}\, z - \frac{t_{0}}{\sqrt{q}}x-\sqrt{\gamma}y}{\sqrt{1-p_1}}\right)}}{{\displaystyle H\left(\frac{\tilde{\kappa}-\sqrt{q}x}{\sqrt{1-q}}\right)}} \,.
	\end{align}
\end{subequations} 
In the previous expression we have redefined $\gamma\equiv p_0-\frac{t_{0}^{2}}{q}$.

As for the RS saddle points analyzed in this paper, we have solved 1RSB equations by a simple iteration scheme. Solving those equations is not a trivial task, mainly because of numerical instabilities but also due to the presence of many integrals in the entropic and energetic terms, which slows a lot the convergence to a fixed-point. 

We have tried to solve these equations for some fixed values of $\alpha$, $\tilde \kappa$, $\kappa$. We have tried several initializations of the order parameter, and different regimes of the overlap $t_1$. For $t_1$ small, i.e. large distances (e.g. in the proximity of the Franz-Parisi entropy) we have not found any solution different from the RS one, i.e. we have always obtained $p_1 = p_0$. This has to be expected. In the small distance regime, due to numerical issues, we were not able to test different initialization for the order parameters; we did not find a non-trivial (i.e. not of the RS type) fixed-point solution to the equations.

\subsection{Numerical validation}

Here we report the details of the numerical experiments shown in Fig.~\ref{Fig::FPentropy} of the main text.

First we collected some sample solutions. We ran the focusing-BP algorithm described in~\cite{unreasoanable} on 10 random samples with $N=2001$ and $P=1000$. We set the number of (virtual) replicas to $y=20$. As a focusing protocol, we performed $30$ steps with the coupling parameter set to $\gamma=\mathrm{atanh}\left(i/29\right)$ where $i=0,\dots,29$ is the step (except for the last step which was capped at $\gamma=10$). At each step, the algorithm was ran to convergence. By the end of the iteration, the distribution was always very strongly peaked on a solution.

For any given pattern set and corresponding solution $\tilde{w}$, we than ran a standard Belief Propagation algorithm with an additional energy term for each variable $w_i$ given by $E_i=-\rho \tilde{w}_i w_i$, i.e. biasing the configurations towards the reference $\tilde{w}$. At convergence, the typical overlap with the reference can be computed as $\frac{1}{N}\sum_i \tilde{w}_i m_i$ where $m_i$ are the magnetizations. Below $\alpha_u$ this overlap is a monotonically increasing function of $\rho$. We let the algorithm converge at $50$ different values of $\rho$, using $\rho=\mathrm{atanh}\left(i/49\right)$ where $i=1,\dots,49$ is the step. For each step, we computed the mean and standard deviations across the 10 samples of both the overlap and the entropy (which can be computed from the Belief Propagation marginals).

\section{The case of non convex neural networks with one hidden layer, general activation functions and binary weights}
We have repeated the same computations (i.e. the typical entropy of solutions and the computation of the local entropy around a typical solution) for a different architecture storing random patters: the tree committee machine. 
It consists in a layer of $K$ non-overlapping perceptrons with $N/K$ weights each, and a second layer of \emph{fixed} weights $c_l$, $l=1, \dots, K$. 
On the output of every of the $K$ perceptrons a point-wise activation function $g(\cdot)$ is also applied. In formulas, given a pattern $\boldsymbol{\xi}$ the corresponding output of the network is
\begin{equation}
	\sigma_{\text{out}} = \text{sign} \left( \frac{1}{\sqrt{K}} \sum_{l=1}^{K} c_l g(x_l)\right)
\end{equation}
where $x_l$ is the preactivation of perceptron $l$:
\begin{equation}
	x_l \equiv \sqrt{\frac{K}{N}} \sum_{i=1}^{N/K} w_{li} \xi_{li} \,.
\end{equation}
In the following we have considered this architecture trained on the same dataset introduced in the main text (i.e. random patterns and labels) and again the case of binary weights ($w_{li} = \pm 1$). The computations are done in the case of generic activation functions (as for example in the paper~\cite{relu_locent} and~\cite{Zavatone2021}), but we consider here for simplicity only the case of the \emph{sign} $g(x) = \text{sign}(x)$ and Rectified Linear Unit (ReLU) $g(x) = \max(0, x)$ activation functions. The weights $c_l$ are all fixed to 1 if we consider the sign activation function and half to +1, half to -1 in the case of the ReLU activation function.
The results we present are all obtained in the limit $N, \, K \to \infty$ with the ratio $K/N \to 0$.
\begin{figure}
	\begin{centering}
		\includegraphics[width=0.75\columnwidth]{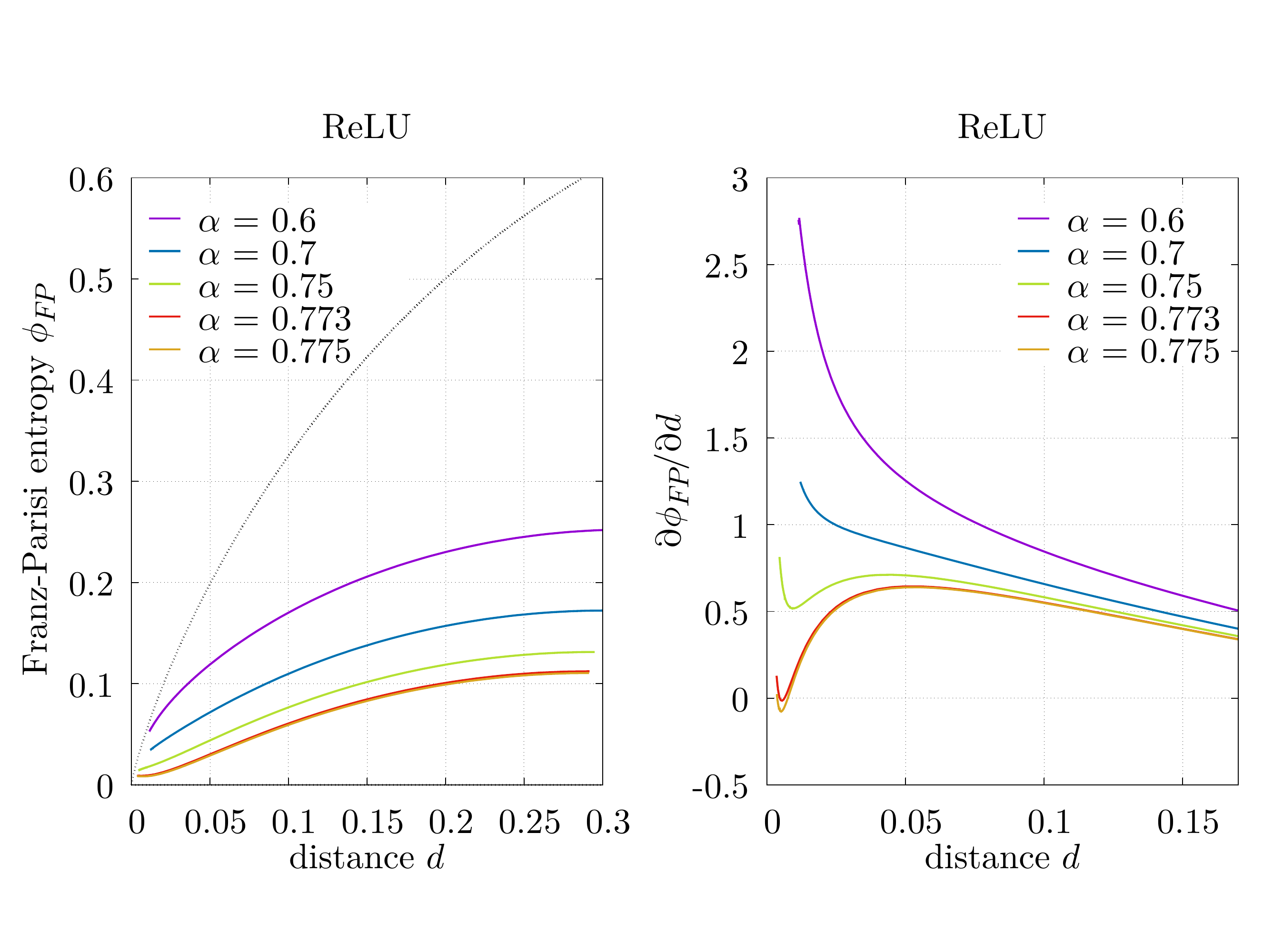}
	\end{centering}
	\caption{\textit{ReLU} activation function: local entropy profiles (with zero margin $\kappa = 0$) of typical maximum margin solutions (left panel) and its derivative (right panel) as a function of the Hamming distance. Different values of $\alpha$ are displayed: for $\alpha=0.6$, $0.7$ and $0.75$ the local entropy is monotonic, i.e. it has a unique maximum at large distances (not visible). For $\alpha = \alpha_{u}^\prime \simeq 0.773$ the local entropy starts to be non-monotonic (its derivative with respect to distance develops a new zero). }
	\label{Fig::FP_treecommittee_relu} 
\end{figure}
\begin{figure}[h]
	\begin{centering}
		\includegraphics[width=0.77\columnwidth]{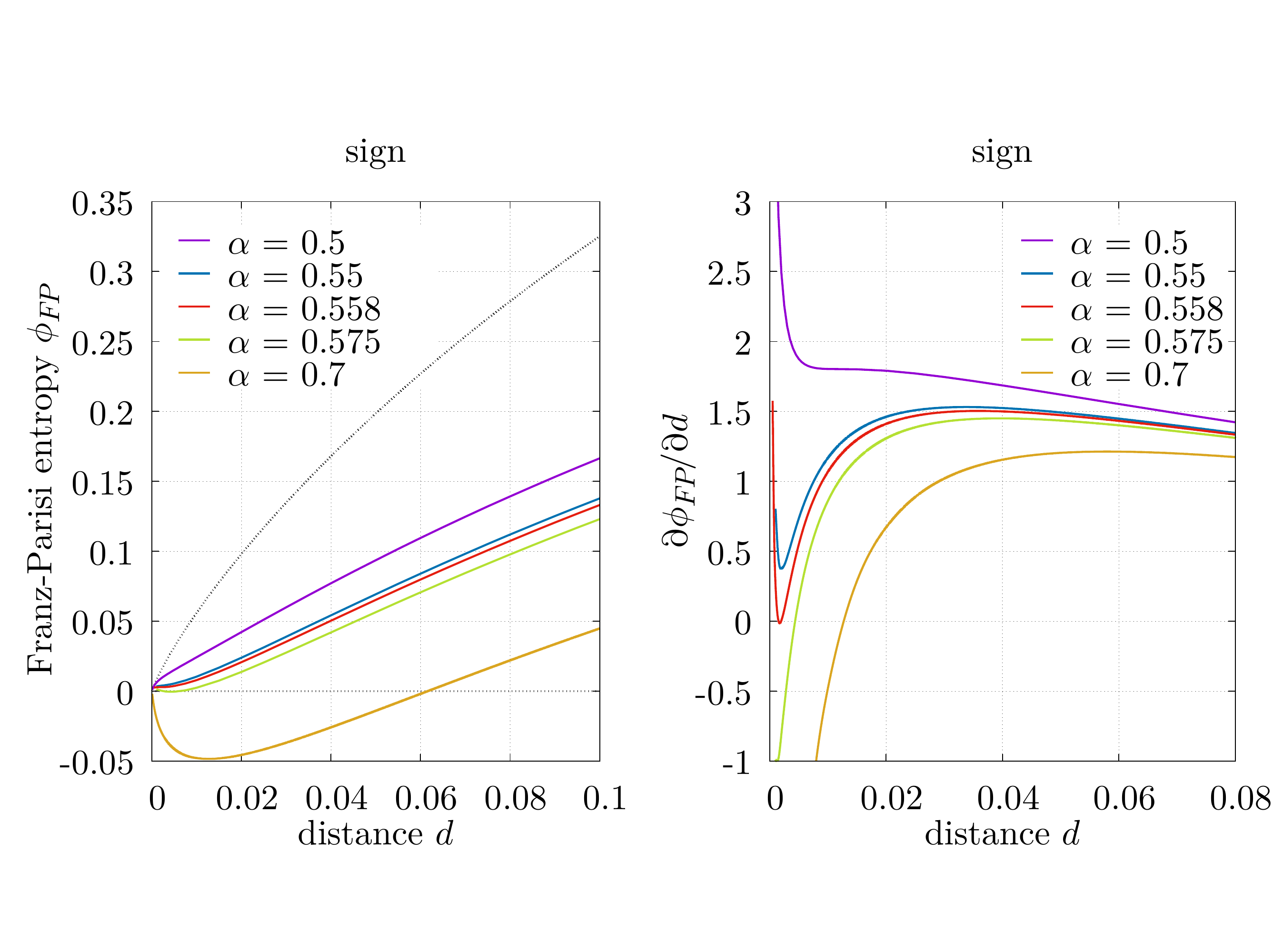}
	\end{centering}
	\caption{\textit{Sign} activation function: local entropy profiles (with zero margin $\kappa = 0$) of typical maximum margin solutions (left panel) and its derivative (right panel) as a function of the Hamming distance. Different values of $\alpha$ are displayed: for $\alpha=0.5$ and $0.55$ the local entropy is monotonic, i.e. it has a unique maximum at large distances (not visible). For $\alpha = \alpha_{u}^\prime \simeq 0.558$ the local entropy starts to be non-monotonic (its derivative with respect to distance develops a new zero). }
	\label{Fig::FP_treecommittee_sign} 
\end{figure}
Concerning the local entropy around typical solutions, we find (in the RS ansatz) that the expression of the entropy is similar to the one obtained with the perceptron architecture~\eqref{eq::FP_entropy}:
\begin{equation}
	\phi_{\text{FP}}=-\frac{\hat{p}}{2}(1-p)+t_{0}\hat{t}_{0}-t_{1}\hat{t}_{1}+\mathcal{G}_{S}+\alpha\mathcal{G}_{E}
\end{equation}
with the same entropic term that we report here for convenience
\begin{eqnarray}
	\mathcal{G}_{S} & = &\int Du\frac{\sum_{\tilde{W}=\pm1}e^{\tilde{W}\sqrt{\hat{q}}x}\int Dv\ln\left[2\cosh\left(\sqrt{\hat{p}-\frac{\hat{t}_0^{2}}{\hat{q}}}v+\frac{\hat{t}_0}{\sqrt{\hat{q}}}u+\left(\hat{t}_1-\hat{t}_0\right)\tilde{W}\right)\right]}{2\cosh\left(\sqrt{\hat{q}}u\right)}
\end{eqnarray}
and an energetic term with ``effective'' order parameters $\Sigma_0$, $\Sigma_1$, $D_0$, $D_1$, $\Delta$, $\Delta_3$:
\begin{eqnarray}
	\mathcal{G}_{E} & = & \int Dx\,\frac{{\displaystyle \int 		Dy\,H\left(\frac{\sqrt{\Gamma}\left(\tilde{\kappa}-\sqrt{\Sigma_0}x\right)- D_1 y}{\sqrt{\Gamma \Sigma_1 - D_1^{2}}}\right)\ln H\left(\frac{\kappa-\frac{D_{0}}{\sqrt{\Sigma_0}}x-\sqrt{\Gamma}y}{\sqrt{\Delta_3}}\right)}}{{\displaystyle H\left(\frac{\tilde{\kappa}-\sqrt{\Sigma_0}x}{\sqrt{\Sigma_1}}\right)}}
\end{eqnarray}
where $\Gamma \equiv \Delta - \frac{D_0^2}{\Sigma_0}$. The effective order parameters depend in particular, on the choice of the activation function and are given by
\begin{subequations}
	\begin{align}
		\Delta &= \int Dx \,  \left( \int Du \, g\left( \sqrt{1-p} u + \sqrt{p} x \right) \right)^2 - \left( \int Dx \, g(x) \right)^2 \\
		\Delta_3 &= \int Dx \left[ g^2(x) - \left( \int Du \, g\left( \sqrt{1-p} u + \sqrt{p} x \right) \right)^2 \right]\\
		\Sigma_0 &= \int Dx \left( \int D\lambda \, g(\sqrt{1-q} \lambda + \sqrt{q} x) \right)^2 - \left( \int Dx \, g(x) \right)^2 \\
		\Sigma_1 &= \int Dx \left[ g^2(x) - \left( \int D\lambda \, g(\sqrt{1-q} \lambda + \sqrt{q} x) \right)^2 \right] \\
		D_0 &= \int Dx D\lambda \, g\left(\lambda \right) \left[ g\left( t_0 \lambda +\sqrt{1-t_0^2} x\right) - g(x) \right]\\
		D_1 &= \int Dx D \lambda \, g\left(\lambda \right) \left[ g\left( t_1 \lambda + \sqrt{1-t_1^2} x \right) - g\left( t_0 \lambda +\sqrt{1-t_0^2} x\right) \right]
	\end{align}
\end{subequations}
All these integrals can be solved in the case of the sign and ReLU activation functions, and we refer to~\cite{relu_locent} for their expressions. 

We show in Fig.~\ref{Fig::FP_treecommittee_relu} and in Fig.~\ref{Fig::FP_treecommittee_sign} the local entropy profiles (with $\kappa = 0$) of the maximum margin solutions for the ReLU and sign activation functions respectively and the corresponding plot of the derivative with respect to Hamming distance. The behavior of the curves for different values of $\alpha$ is similar to the picture presented in the main text for the simpler perceptron architecture. In particular we find that the monotonic to non-monotonic transition is $\alpha_u^\prime\simeq 0.773$ for the ReLU activation whereas we found a much lower value $\alpha_u^\prime \simeq 0.558$ for the sign activation function.

\section{Numerical results on deep networks}

Here we show that our results are consistent with experiments on deep multi-layer perceptrons (MLP) trained on different architectures and different real-world datasets.
In particular we vary the size of the dataset and for each size we train the networks repeatedly, starting from different random initial conditions, and measure the distances between the final configurations. We find that the final configurations are always far apart. When the number of patterns is sufficiently low the train error is zero, up to the algorithmic threshold where the dataset size becomes too large for the network to memorize perfectly without errors. Even then, different solutions are still far apart, thus implying that many solutions exist even at the algorithmic threshold.

We present results on three different architectures trained on different datasets: a two hidden layer MLP with width $101$ trained on odd-even digits of MNIST, a two hidden layer MLP with width $201$ trained on FashionMNIST and a three hidden layer MLP with width $501$ trained on CIFAR10.

In order to train a binary weights network we used a standard BinaryNet implementation \cite{hubara2016binarized} without biases and batch normalization.
In all the experiments we used the cross-entropy loss and trained the model with SGD optimization (lr=$5$ for MNIST and FashionMNIST and lr=$10$ for CIFAR10) for a fixed number of epochs ($200$ epochs for all the experiments).
For each number of pattern, we collected different solutions by training the models with different random initializations ($20$ samples for MNIST and FashionMNIST and $15$ samples for CIFAR10).

Before computing the average overlap between pairs of solutions, we need to take into account the symmetries in the network, namely the permutation symmetry among neurons belonging to the same layer, and the sign reversal symmetry. In practice, for each pair of solutions we perform a matching (taking the sign of the neuron's weights into account) on each layer, starting from the bottom one and adjusting accordingly the neurons of the next layer, in order to break the symmetries, and then compute their overlaps (see also \cite{baldassi2021learning}). This procedure guarantees that two networks that differed only because of the labeling of their hidden units or because of sign reversals would still be identified as the same network and would have overlap $1$.

In order to assess the flatness of the minima at the algorithmic threshold, we computed the local energy profiles (see \cite{pittorino2020entropic} for a discussion on the relation between local entropy and local energy) of solutions at the first value of the number of patterns at which the average train error is exactly zero ($P=1521$ for MNIST, $P=100$ for FashionMNIST and $P=8643$ for CIFAR10). 
Given a solution, we perturb it by flipping a random fraction of the weights to obtain the number of errors at a given distance. Varying the flip probability and averaging over several realization of the noise ($20$ noise realizations for the experiments reported here), we obtain the curves for the error as a function of the distance from the reference solution.

\begin{figure}[h]
	\begin{centering}
		\includegraphics[width=0.3\columnwidth]{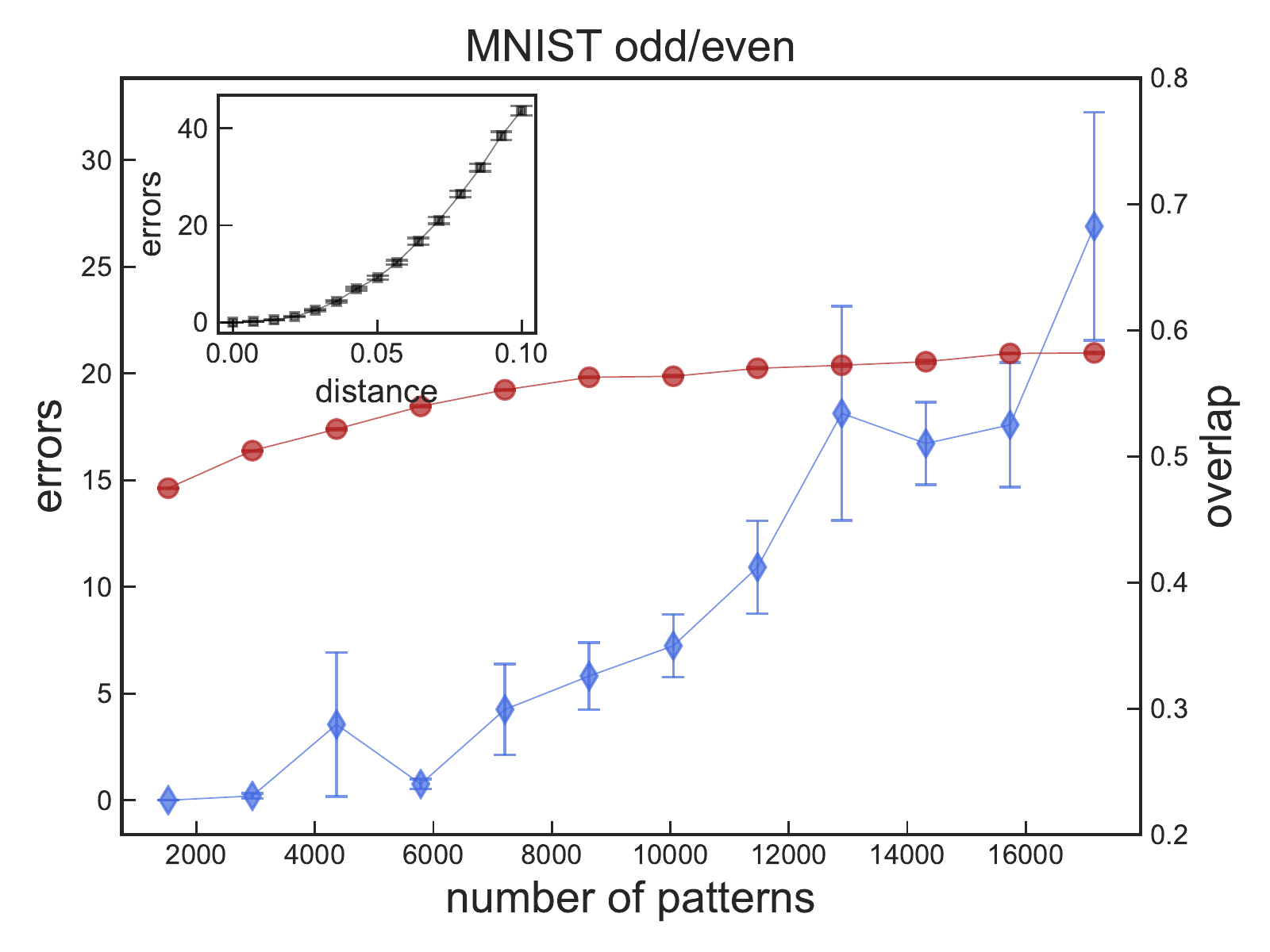}
		\includegraphics[width=0.3\columnwidth]{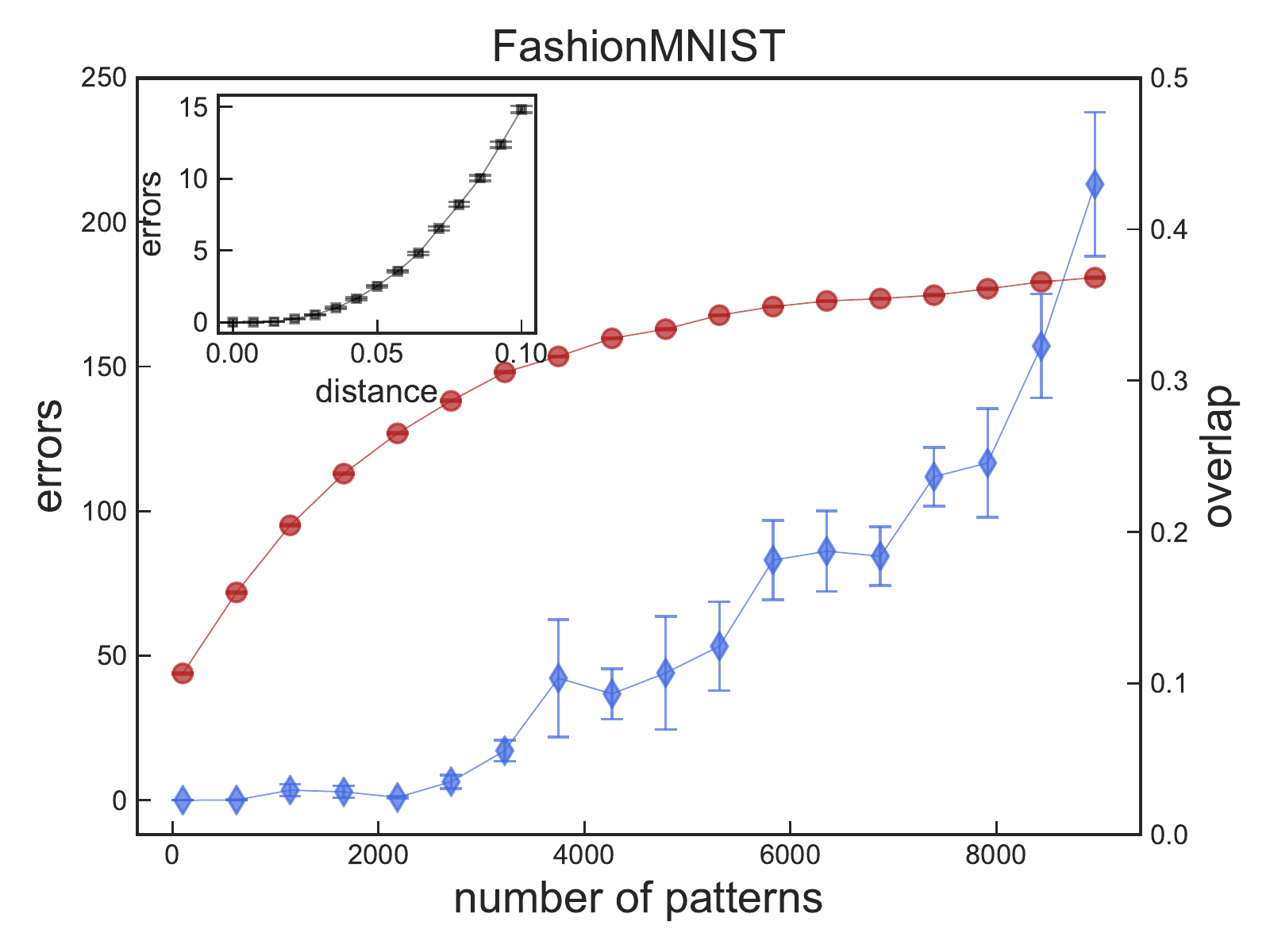}
		\includegraphics[width=0.3\columnwidth]{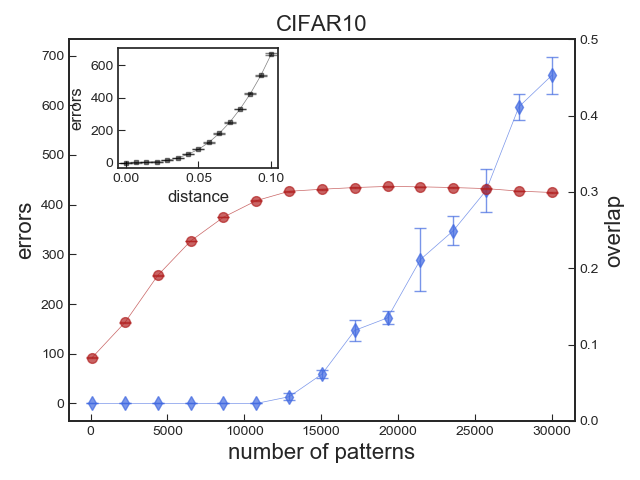}
	\end{centering}
	\caption{Number of train errors (blue points, left y-axis) and mean overlap of pairs of solutions (red points, right y-axis) as a function of the training set size, for different architectures and datasets. From left to right: binary classification of odd/even digits of MNIST, FashionMNIST, and CIFAR10. The overlaps between solutions obtained with different random initializations is well below $1$ even as the model is barely able to fit all of the patterns. The inset shows local energy profiles (see text) right below the algorithmic threshold, i.e. at the largest dataset size for which the error is still zero.}
	\label{Fig::MLPs} 
\end{figure}

The results are reported in Fig.~\ref{Fig::MLPs}. The scenario is consistent among all the architectures and the datasets tested, and shows that many solutions exist at the algorithmic threshold as the average overlap is always well below $1$. Moreover, when the number of parameters is such that the model is able to perfectly fit all the patterns, the solutions found are not isolated, as their local energy profile has the typical convex shape of flat minima (see insets in Fig.~\ref{Fig::MLPs}). 

\end{document}